\documentclass[
]{jss}

\usepackage[utf8]{inputenc}

\providecommand{\tightlist}{%
  \setlength{\itemsep}{0pt}\setlength{\parskip}{0pt}}

\author{
Nicholas Tierney\\Monash University \And Dianne Cook\\Monsh University
}
\title{Expanding tidy data principles to facilitate missing data exploration, visualization and assessment of imputations}

\Plainauthor{Nicholas Tierney, Dianne Cook}
\Plaintitle{Expanding tidy data principles to facilitate missing data exploration, visualization and assessment of imputations}
\Shorttitle{Explore missingness with \pkg{naniar}}

\Abstract{
Despite the large body of research on missing value distributions and imputation, there is comparatively little literature with a focus on how to make it easy to handle, explore, and impute missing values in data. This paper addresses this gap. The new methodology builds upon tidy data principles, with the goal of integrating missing value handling as a key part of data analysis workflows. We define a new data structure, and a suite of new operations. Together, these provide a connected framework for handling, exploring, and imputing missing values. These methods are available in the R package \texttt{naniar}.
}

\Plainkeywords{statistical computing, statistical graphics, data science, data visualization, tidyverse, data pipeline, R}

\Address{
    Nicholas Tierney\\
  Monash University\\
  Department of Econometrics and Business Statistics E774, Menzies Building Monash University, Clayton\\
  E-mail: \email{nicholas.tierney@gmail.com}\\
  URL: \url{http://www.njtierney.com}\\~\\
    }

\usepackage{color}
\usepackage{fancyvrb}

\DefineVerbatimEnvironment{Highlighting}{Verbatim}{commandchars=\\\{\}}
\usepackage{framed}
\definecolor{shadecolor}{RGB}{248,248,248}

\usepackage{amsmath} \usepackage[english]{babel}

\begin{document}

\hypertarget{intro}{%
\section{Introduction}\label{intro}}

As data science has become a more solid field, theories and principles have developed to describe best practices. One such idea is `tidy data,' which defines a clean, analysis-ready format that informs workflows converting raw data through a data analysis pipeline \citep{Wickham2014}. Another idea is the grammar of graphics, describing how to map data values into a visualization \citep{Wilkinson2012}. These principles have been widely adopted, but do not address the problem of missing data. In particular, there is little guidance on how to handle missing values in a data analysis workflow. Most analysis and visualization software simply drop missing values when making a plot, although some (\pkg{ggplot2}) provide a warning (Figure \ref{fig:warning}).

\begin{CodeChunk}
\begin{figure}

{\centering \includegraphics[width=0.75\linewidth]{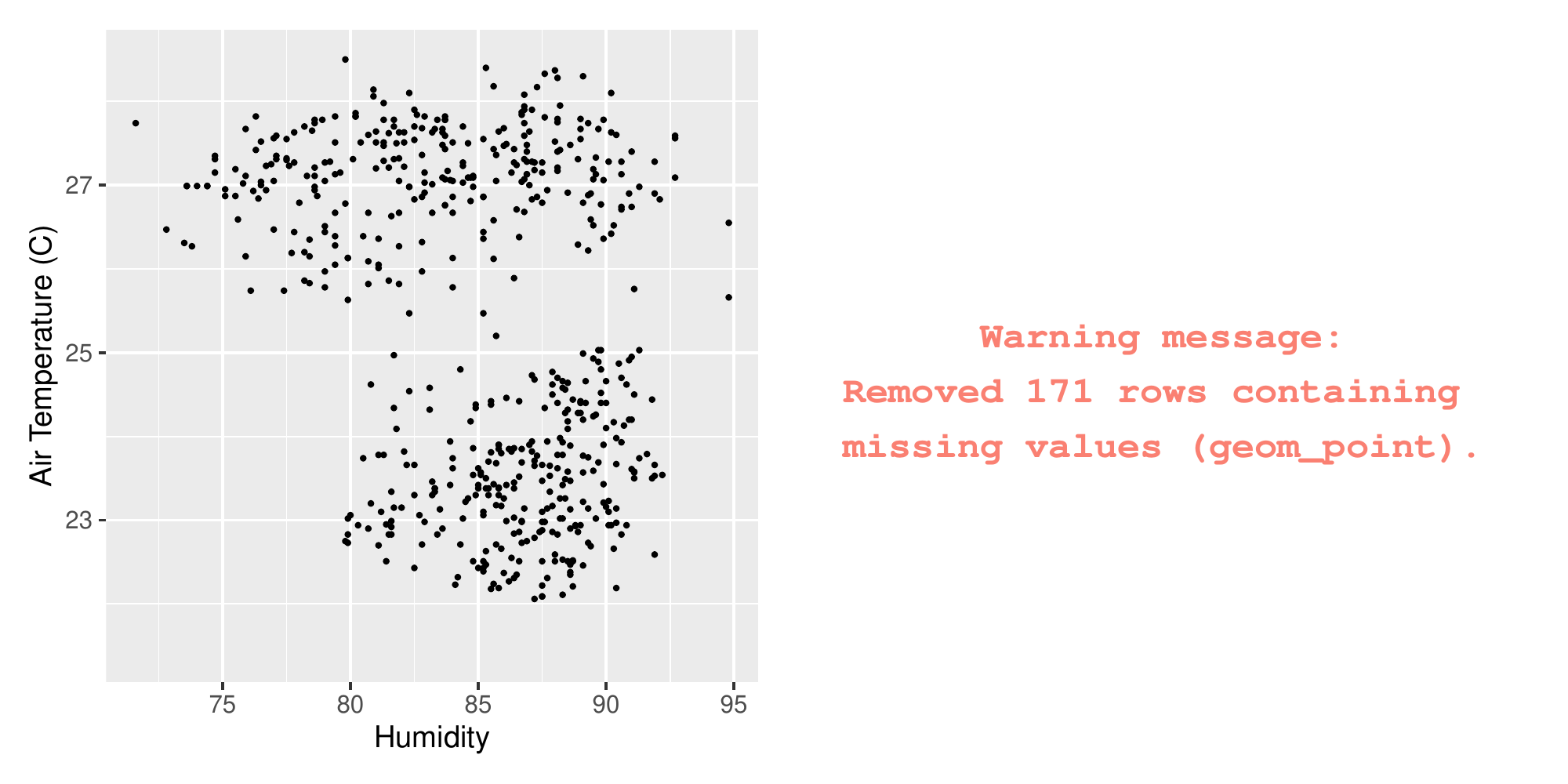} 

}

\caption[How ggplot2 behaves when displaying missing values]{How ggplot2 behaves when displaying missing values. A warning message is displayed, but missing values are not shown in the plot}\label{fig:warning}
\end{figure}
\end{CodeChunk}

The imputation literature focuses on ensuring valid statistical inference
is made from incomplete data. This is approached chiefly through probabilistic
modelling, assuming the mechanism of missing data is known to the analyst. It
does not address how to explore, understand, and handle missing data structures
and mechanisms.

However, something must be known about the missing value structure to produce a
complete dataset for analysis, whether by case- or variable-deletion, or with
imputation. Decision tree models or latent group analysis can reveal structures or patterns
of missingness \citep{Tierney2015, Barnett2017}, but are not definitive. While there
are times where the missing data mechanism is obvious, determining this is typically not straightforward. To understand their structure and possible mechanisms, the analyst must explore the data
with visualizations, summaries, and modelling, in an iterative fashion.

While there have been many software tools for exploring missing data, they do not work together. The graphics literature provides several solutions for exploring missings visually by incorporating them into the plot in some way. For example, imputing values to be 10\% below the minimum to include all observations into a scatter plot \citep{Cook2007}, or treating missing values as an equal category of data \citep{Unwin1996}. The ideas from the graphics literature need to be translated into tidy data tools to integrate missing data handling in a data analysis pipeline.

This paper is organized in the following way. The next section (\ref{background}) provides the background to tidy data principles and tools (\ref{tidy-data-concepts}) and missing data representations (\ref{missing-data-rep-dep}). Section \ref{existing-software} summarises existing software for handling missing data. The new work extending the tidy tools to facilitate exploring, visualizing, and imputing missing data is discussed in Section \ref{extensions}. Graphics (\ref{graphics}) and Numerical summaries (\ref{num-sum}) of missing values are then discussed. An application illustrating the use of the new methods is shown in Section \ref{case-study}. Section \ref{discussion} discusses strengths, limitations, and future directions.

\hypertarget{background}{%
\section{Background}\label{background}}

\hypertarget{tidy-data-concepts}{%
\subsection{Tidy data concepts and methods}\label{tidy-data-concepts}}

Features of tidy data were formally described in \citet{Wickham2014}, and were discussed in terms of their importance for data science by \citet{Donoho2017}, and tools for data analysis. Tidy data is defined by \citet{r4ds} as:

\begin{quote}
\begin{enumerate}
\def\labelenumi{\arabic{enumi}.}
\tightlist
\item
  Each variable must have its own column.
\item
  Each observation must have its own row.
\item
  Each value must have its own cell.
\end{enumerate}
\end{quote}

Tidy data is easier to work with and analyze because the variables are in the same format as they would be put into modelling software. This helps the analyst works swiftly and clearly, closing up opportunities for errors. Tidy data principles are general, but are comprehensively implemented in the R programming language, so this paper focuses on R \citep{rcore}.

Tidy tools require the same tidy data input and output. This consistency means multiple tools can be composed together into a sequence, allowing for rapid, elegant, and complex operations. Contrasting tidy tools are messy tools. These have tidy input but messy output. Messy tools slow down analysis by shifting the focus from analysis to transforming output so it is the right shape for the next step in the analysis. This makes the work at each step harder to predict, and more complex and difficult to maintain. This disrupts workflow, and invites errors. Tidy tools fall into three broad categories: data manipulation, visualization, and modelling.

\hypertarget{tidy-data-manip}{%
\subsubsection{Data manipulation}\label{tidy-data-manip}}

Data manipulation is made input- and output-tidy with R packages \pkg{dplyr} and \pkg{tidyr} \citep{dplyr, tidyr}. These provide the five ``verbs'' of data manipulation: data reshaping, sorting, filtering, transforming, and aggregating. Data reshaping goes from long to wide formats; sorting arranges rows in a specific order; filtering removes rows based on a condition; transforming, changes existing variables or adds new ones; aggregating creates a single value from many values, say, for example, in computing the minimum, maximum, and mean.

\hypertarget{tidy-vis}{%
\subsubsection{Visualizations}\label{tidy-vis}}

Visualization tools only have tidy data as their input, as the output is a graphic. The popular domain specific language \pkg{ggplot2} maps variables in a dataset to features (referred to as aesthetics) of a graphic \citep{ggplot2}. For example, a scatterplot can be created by mapping two variables to the x and y axes, and specifying a point geometry.

\hypertarget{tidy-model}{%
\subsubsection{Modelling}\label{tidy-model}}

Modelling tools work well with tidy data, as they have a clear mapping from variables in the data to the formula for a model. For example in R, y regressed on x and z is: \code{lm(y ~ x + z)}. Modelling tools are input tidy, but their output is always messy - it is not in the right format for subsequent steps in analysis. For example, estimated coefficients, predictions, and residuals from one model cannot be easily combined with the output of another model. Messy models have been partially addressed with the \pkg{broom} package, which tidies up model outputs into a tidy data format for data analysis, and the developing \pkg{recipes} package, which helps make modelling input- and output-tidy \citep{recipes}.

\hypertarget{the-tidyverse}{%
\subsubsection{The tidyverse}\label{the-tidyverse}}

Defining tidy data and tidy tools has resulted in a growing set of packages known collectively as the ``tidyverse'' \citep{tidyverse}. These are constructed to share similar principles in their design and behavior, and cover the breadth of an analysis - from importing, tidying, transforming, visualizing, modelling, to communicating \citep{r4ds, tidyverse, Tidyverse-Manifesto}. This has led to more tools for specific parts of analysis - from reading in data with \pkg{readr}, \pkg{readxl}, and \pkg{haven}, to handling character strings with \pkg{stringr}, dates with \pkg{lubridate}, and performing functional programming with \pkg{purrr} \citep{readr, readxl, haven, stringr, lubridate, purrr}. It has also led to a burgeoning of new packages for other fields following similar design principles, creating fluid workflows for new domains. For example, the \pkg{tidytext} \citep{tidytext} package for text analysis, the \pkg{tsibble} \citep{wang2020tsibble} package for time series data, and \pkg{tidycensus} \citep{tidycensus} for working with US census and boundary data.

\hypertarget{tidy-formats-missing-data}{%
\subsubsection{Tidy formats for missing data}\label{tidy-formats-missing-data}}

Current tools for missing data are messy. Missing data tools can be used to perform imputations, missing data diagnostics, and data visualizations. However, these tools suffer the same problems as modelling: They use tidy input, but produce messy output - their output is challenging to integrate with other steps of data analysis. The complex, often multivariate nature of imputation methods also makes makes them difficult to represent. Visualization methods for missing data do not map data features to the aesthetics of a graphic, as in \pkg{ggplot2}, limiting expressive exploration.

Taking existing methods from the missing data graphics literature, and translating and expanding them into tidy data and tidy tools would create more effective data visualizations.
Defining these concepts allows the focus to be more general than just software, but rather, an extensible framework for tidy tools to explore missing data.

\hypertarget{missing-data-rep-dep}{%
\subsection{Missing data representation and dependence}\label{missing-data-rep-dep}}

The convention for representing missingness is a \textbf{b}inary matrix, \(B\), for data \(y\) with \(i\) rows and \(j\) columns:

\[
b_{ij} =\begin{cases}
1, & \text{if } y_{ij} \text{ is missing} \\
0, & \text{if } y_{ij} \text{ is observed}
\end{cases}
\]

There are many ways each value can be missing, we adopt the notation used in \citet{VanBuuren2012}. The information in \(B\) can be used to arrive at three categories of missing values: Missing completely at random (MCAR), missing at random (MAR), and missing not at random (MNAR). The distribution of missing values in \(b_{ij}\) can depend on the entire dataset, represented as \(Y = (Y_{obs}, Y_{miss})\). This relationship can be defined by the \emph{missing data model} \(Pr(b_{ij} | Y_{obs}, Y_{miss}, \psi)\), the probability of missingness is conditional on data observed, data missing, and some probability parameter of missingness, \(\psi\). This helps to precisely define categories of missing values.

\textbf{MCAR} is where values being missing have no association with observed or unobserved data, that is, \(Pr(B = 1 | Y_{obs}, Y_{miss}) = Pr(B = 1 | \psi)\). Essentially, the probability of an observation being missing is unrelated to anything else, only the parameter \(\psi\), the overall probability of missingness. Although a convenient scenario, it is not actually possible to confirm, or clearly distinguish from MAR, as it relies on statements on data unobserved. In \textbf{MAR}, missingness only depends on data observed, not data missing, that is, \(Pr(B = 1 | Y_{obs}, Y_{miss}, \psi) = Pr(B = 1 | Y_{obs}, \psi)\). Some structure or dependence between missing and observed values is allowed, provided it can be explained by data observed, and some overall probability of missingness. In \textbf{MNAR}, missingness is related to values observed, and unobserved: \(Pr(B = 1 | Y_{obs}, Y_{miss}, \psi)\). This assumes conditioning on all observations: data goes missing due to some phenomena unobserved, including the structure of the missing data itself. This presents a challenge in analysis, as it is difficult to verify, and implies bias in analysis due to the unobserved phenomena. Visualizations can help assess whether data may be MCAR, MAR or MNAR.

\hypertarget{existing-software}{%
\section{Existing Software}\label{existing-software}}

Methods for exploring, understanding, and imputing missing data are more accessible now than they have ever been. Values can be imputed with one value (single imputation), or multiple values (multiple imputation), creating \(m\) datasets. This section discusses existing software for single and multiple imputation, and missing data exploration.

\hypertarget{imputation}{%
\subsection{Imputation}\label{imputation}}

\pkg{VIM} \citep{VIM} implements well-used imputation methods K nearest neighbors, regression, hot-deck, and iterative robust model-based imputation. These diverse approaches allows for imputing with semi-continuous, continuous, count, and categorical data. VIM identifies imputed cases by adding an indicator variable with a suffix \texttt{\_imp}. So \texttt{Var1} has a sibling column, \texttt{Var1\_imp}, with values TRUE or FALSE indicating imputation. \pkg{VIM} also has a variety of visualization methods, discussed in \ref{exploration}. \pkg{simputation} provides an interface to imputation methods from VIM, in addition providing hotdeck imputation, and the EM algorithm \citep{simputation, Dempster1977}. \pkg{simputation} provides a consistent formula interface for all imputation methods, and always returns a dataframe with the updated imputations. \pkg{Hmisc} \citep{Hmisc}, provides predictive mean matching, \pkg{imputeTS} \citep{imputeTS}, provides time series imputation methods, and \pkg{missMDA} \citep{missMDA}, imputes data using principal components analysis.

Multiple imputation is often regarded as best practice for imputing values \citep{Schafer2002}, as long as appropriate caution is taken \citep{Sterne2009}. Popular and robust methods for multiple imputation include the \pkg{mice}, \pkg{Amelia}, and \pkg{mi} packages \citep{mice, amelia, mi}. \pkg{mice} implements the method of chained equations, using a variable-wise algorithm to calculate the posterior distribution of parameters to generate imputed values. The workflow in \pkg{mice} revolves around imputing data, returning completed data, and fitting a model and pooling the results.

\pkg{Amelia} \citep{amelia} assumes data are multivariate normal, and samples from the posterior, and allows for incorporation of information on the values in a prior. It uses the computationally efficient (and parallelizable) Expectation-Maximization Bootstrap (EMB) algorithm \citep{Honaker2010}. \pkg{norm} \citep{norm}, provides multiple imputation using EM for multivariate normal data, drawing from methods in the NORM software \citep{schafer-norm}. \pkg{norm} does not provide a framework for tracking missing values, instead providing tools for making inference from multiple imputation.

\pkg{mi} \citep{mi} also uses Bayesian models for imputation, providing better handling of semi-continuous values, and data with structural or perfect correlation. A collection of analysis models are also provided in \pkg{mi}, to work with data it has imputed. These include linear models, generalized linear models, and their corresponding Bayesian components. This approach promotes fluid workflow, with a similar penalty to tidying up model output, which is still messy.

\hypertarget{imputation-summary}{%
\subsubsection{Summary}\label{imputation-summary}}

Each imputation method provides practical methods for different use cases, but most have different output structures, and do not have consistent interfaces in their implementation. This makes them inherently messy and challenging to integrate into an analysis pipeline. For example, combining different imputation methods from different pieces of software is not currently straightforward. \pkg{simputation} resolves some of these complications with a simple approach of a unified syntax for all imputation, and always returns a dataframe of imputed values. This reduces the friction of working with other tools, but comes at the cost of identifying imputed values. An ideal approach would use consistent, simple data structures that work with other analysis tools, and help track missing values. This would make imputation outputs tidy, streamlining subsequent analysis.

\hypertarget{exploration}{%
\subsection{Exploration}\label{exploration}}

The primary focus of most missing data packages is making inferences, and exploring imputed values, not
on exploring relationships in missing values, and identifying possible patterns.
Texts covering the exploration phase of missing data have the same problem as with modelling: the input is tidy, but the output does not work with other tools \citep{VanBuuren2012}; this is inefficient. Methods for exploring missing values are primarily covered in literature on interactive graphics \citep{Swayne1998, Unwin1996, Cook2007}, and are picked up again in a discussion of a graphical user interface \citep{Cheng2015}.

The missingness matrix \(B\) can be used to assess missing data dependence. It has been used in interactive graphics, dubbed a ``shadow matrix'', to link missing and imputed values to the data, facilitating their display \citep{Swayne1998}, focusing heavily on multivariate numeric data. This is an idea upon which this new work builds.

The MANET (Missings Are Now Equally Treated) software \citet{Unwin1996} focussed on multivariate categorical data, with missingness explicitly added as a category. MANET also provided univariate visualizations of missing data using linked brushing between a reference plot of the missingness for each variable, and a plot of the data as a histogram or barplot. The MANET software is no longer maintained and cannot be installed. The approach of \citet{Swayne1998} in the software XGobi, further developed in ggobi \citep{Cook2007}, focussed on multivariate quantitative data. Missingness is incorporated into plots in ggobi by setting them to be 10\% below the minimum value.

MissingDataGUI provides a Graphical User Interface (GUI) for exploring missing data structure, both numerically and visually. Using a GUI to explore missing data facilitates rapid insight into missingness structures. However, this comes as a trade off, as insights are not captured or recorded with a GUI, making it challenging to incorporate into reproducible analyses. This distracts and breaks analysis workflow, inviting mistakes.

VIM (Visualizing and Imputing Missing Data) provides visualization methods to identify and explore observed, imputed, and missing values. These include spinograms, spinoplots, missingness matrices, plotting missingness in the margins of other plots, and other summaries. However, these visualizations do not map variables to graphical aesthetics, creating friction when moving through analysis workflows, making them difficult to extend to new circumstances. Additionally, data used to create the visualizations cannot be accessed, posing a barrier to further exploration.

\pkg{ggplot2} removes missing values with a warning (Figure \ref{fig:warning}), and only incorporates missingness into visualizations when mapping a discrete variable with missings to a graph aesthetic. This has some limitations, shown in Figure \ref{fig:gg-box-na}, a boxplot visualization of school grade and test scores. If there are missings in a continuous variable like test score, \pkg{ggplot2} omits the missings and prints a warning message. However, if a discrete variable like school year has missing values, an NA category is created for school year, where scores are placed (Figure \ref{fig:gg-box-na}).

\begin{CodeChunk}
\begin{figure}

{\centering \includegraphics[width=1\linewidth]{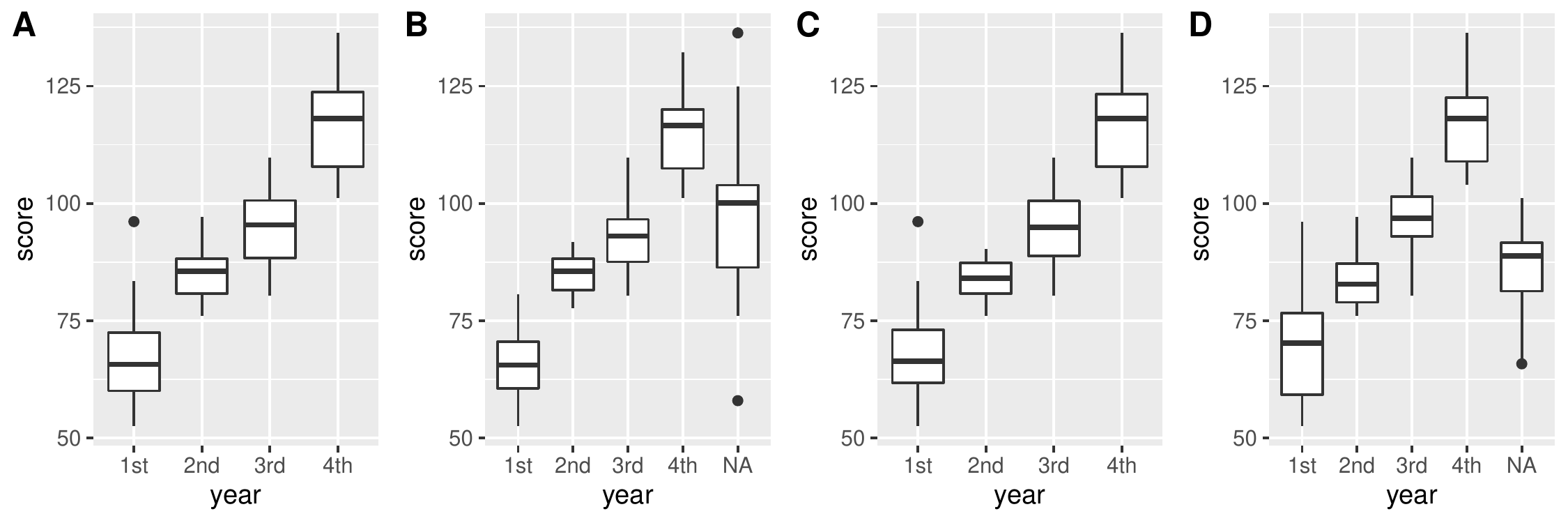} 

}

\caption[ggplot2 provides different visualizations depending on what type of data has missing values for data of student test scores in school year]{ggplot2 provides different visualizations depending on what type of data has missing values for data of student test scores in school year. (A) Data is complete; (B) Missings are only in year - an additional 'NA' boxplot is created; (C) Missings only in scores, no additional missingness information is shown; (D) Missings in both scores and year, additional missing information is shown. The missingness category is only shown when there are missings in categorical variables such as year (plots (B) and (D)). In (C), no missingness information is given on the graphic, despite there being missings in score, and a warning message is displayed about the number of missing values omitted.}\label{fig:gg-box-na}
\end{figure}
\end{CodeChunk}

\hypertarget{extensions}{%
\section{Tidy framework for missings}\label{extensions}}

Applying tidyverse principles has the potential to clarify missing data exploration, visualization, and imputation. This
section discusses how these principles are applied for data structures (\ref{data-structure}), common operations (verbs) (\ref{verbs}), graphics (\ref{graphics}), and data summaries (\ref{num-sum}). Care has been taken to make the names and design these features intuitive for their purpose, and is discussed throughout.

\hypertarget{data-structure}{%
\subsection{Data structure}\label{data-structure}}

A data structure facilitating exploration of missing data needs to be simple to reason with and to transport with a data analysis, otherwise it will not be used. A useful template common in missing data literature, is the \(B\) matrix, where 0 and 1 indicate not missing and missing, respectively. This matrix was used to explore missing values in the interactive graphics library XGobi, called a ``missing value shadow'', or ``shadow matrix'', defined as a copy of the original data with
indicator values of missingness. The shadow matrix could be interactively linked to the data. However, there are some limitations to the shadow matrix. Namely, the values 0 and 1 can be confusing representations of missing values, since it is not clear if 0 indicates an absence of observation, or the presence of a missing value.

We propose a new form for tidy representation of missing data based on these ideas from past research. Four features are added to the shadow matrix to facilitate analysis of missing data, illustrated in Figure \ref{fig:nabularfig}.

\begin{enumerate}
\def\labelenumi{\arabic{enumi}.}
\item
  \textbf{Missing value labels}. Simple labels for missing and not missing to clearly identify missing values for analysis and plotting. We propose ``NA'' and ``!NA''. (Figure \ref{fig:nabularfig}). This improves the 0 and 1 values in \(B\), which do not clearly identify missingness. These values follow a principle: ``\textbf{clarity of labelling}'' - the matrix's meaning is transparent, and anybody looking at these values could understand what they mean, which is not the case of binary values. Equally, these values could instead be ``missing'' or ``present''.
\item
  \textbf{Special missing values}: Building on \textbf{missing value labels}, the values in the shadow matrix can be ``special'' missing values, indicated by ``NA\_\textless suffix\textgreater{}''. For example: \texttt{NA\_instrument} uses a short label, ``instrument'', indicating instrument error resulting in missing values. These could be also used to indicate imputations, (Figure \ref{fig:nabularfig}).
\item
  \textbf{Coordinated names}: Variable names in the shadow matrix gain a consistent short suffix, "\_NA", keeping names coordinated throughout analysis (Figure \ref{fig:nabularfig}). It makes a clear distinction with \texttt{var\_NA} being a random variable of the missingness of a variable, \texttt{var}. This suffix is short and easy to remember during data analysis, and shifts the focus from the value of a variable, to its missingness state.
\item
  \textbf{Connectedness}: Binding the shadow matrix column-wise to the original data creates a single, connected, \emph{nabular} data format, in sync with the data. It is useful for visualization, summaries, and tracking imputed values, discussed in more detail in \ref{nabular-data}.
\end{enumerate}

\hypertarget{nabular-data}{%
\subsection{Nabular data}\label{nabular-data}}

\emph{Nabular} data binds the shadow matrix column-wise to the original data. It is a portmanteau of \texttt{NA} and \texttt{tabular}. \emph{Nabular} data explicitly links missing values to data, keeping corresponding observations together and removing the possibility of mismatching records (Figure \ref{fig:nabularfig}). \emph{Nabular} data facilitates visualization and summaries by allowing the user to reference the missingness of a variable in a coordinated way: the missingness of \texttt{var}, as \texttt{var\_NA} during analysis. \emph{Nabular} data is a snapshot of the missingness of the data. This means when \emph{nabular} data are imputed, those imputed values can easily be identified in analysis (\texttt{var} vs \texttt{var\_NA}). \emph{Nabular} data is not unlike classical data formats with quality or flag variables associated with each measured variable, e.g.~Scripps CO2 data \citep{Keeling2005-scripps}, GHCN data \citep{Durre2008-ghcn}.

\begin{CodeChunk}
\begin{figure}

{\centering \includegraphics[width=1\linewidth]{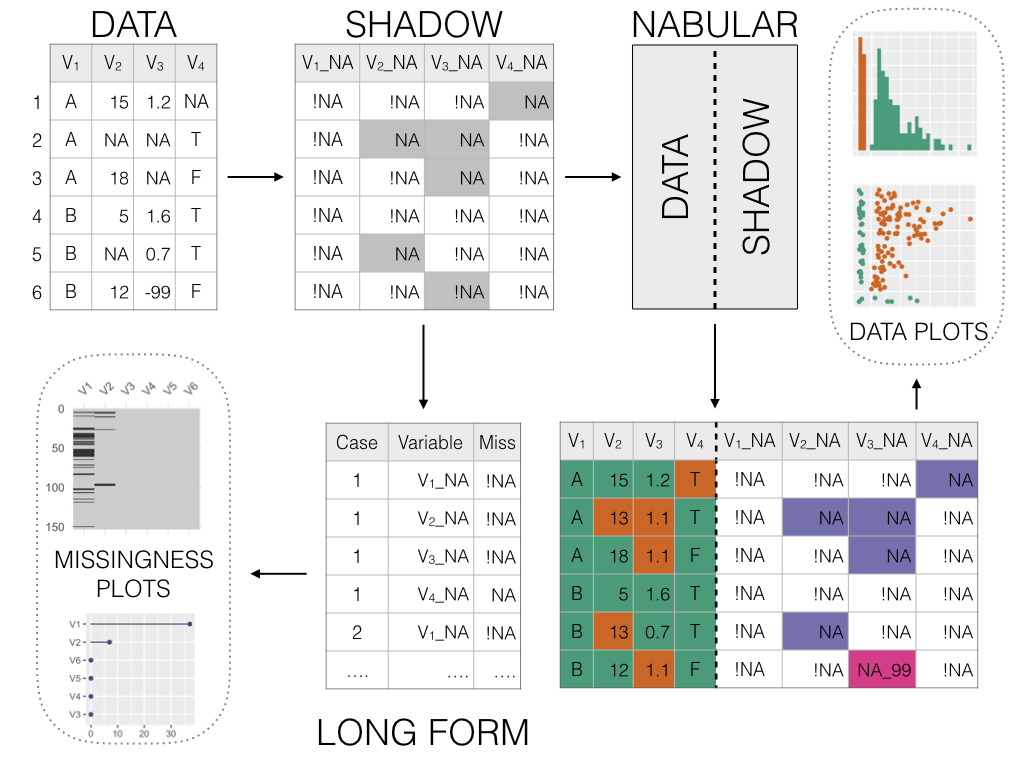} 

}

\caption[The process of creating nabular data]{The process of creating nabular data. Data transformed to shadow matrix, where values are either not missing or missing: '!NA' or 'NA'. The shadow matrix can be converted to long form to create missingness summary plots. Nabular data is created by binding the columns of the data and shadow matrix. Special missing values (such as -99) are identified as special missings, and values imputed and tracked. Nabular data can be used to identify imputations and explore data values alongside missings, providing a useful format for missing data exploration and analysis.}\label{fig:nabularfig}
\end{figure}
\end{CodeChunk}

Although a binary missingness matrix, \(B\), could be generated during analysis instead of using a \emph{nabular} data structure, there are key advantages to using \emph{nabular} data. Firstly, missing value labels \texttt{NA} and \texttt{!NA} are clearer than TRUE or FALSE. Secondly, special missing values cannot be added easily during analysis with a logical matrix. Finally, the logical matrix cannot capture which values are imputed if imputation has already taken place. Imputing values on \emph{nabular} data automatically tracks these imputations.

Using additional columns to represent missingness information follows best practices for data organization, described in \citet{Ellis2017} and \citet{Broman2017}: (1) Keep one thing in a cell and (2) Describe additional features of variables in a second column. Here they suggest to indicate censored data with an extra variable called ``VariableNameCensored'', which would be TRUE if censored, otherwise FALSE. This information can now be represented in the shadow columns as special missing values. Encoding special missing values is achieved by defining logical conditions and suffixes. This is implemented with the \texttt{recode\_shadow} function in \pkg{naniar} (\ref{verbs-recode}).

Special missing values are not a new idea, and have been implemented in other statistical programming languages, SPSS, SAS, and STATA. These typically represent missing values as a full-stop, \texttt{.}, and record special missing values as \texttt{.a} - \texttt{.z}. These special values from these languages break the tidy principle of one column having one type of value, as they record both the value, and the multivariate missingness state.

\hypertarget{verbs}{%
\subsection{Missing data operations}\label{verbs}}

Common missing data operations can be considered verbs, in the tidyverse sense. For missing data, these include: \textbf{scan}, \textbf{replace}, \textbf{add}, \textbf{shadow}, \textbf{impute}, \textbf{track}, and \textbf{flag}. Data can be \textbf{scanned} to find possible missings not coded as \texttt{NA}. These values can then be \textbf{replaced} with \texttt{NA}. To facilitate exploration, summaries of missingness can be \textbf{added} as a summary column to the original data. The data can be augmented with the \textbf{shadow matrix} values, helping explore missing data, as well as facilitating the process of \textbf{imputing}, and \textbf{tracking}. Finally, unusual or specially coded missing values can be \textbf{flagged}.

\hypertarget{verbs-search}{%
\subsubsection{scan: Searching for common missing value labels}\label{verbs-search}}

This operation is used to search the data for specific conventional representations of missings, such as, ``N/A'', ``MISSING'', \texttt{-99}.
This is implemented in the function \texttt{miss\_scan\_count()}, which returns a table of occurrences of that value for each variable. A list of common NA values for numbers and characters, can be provided to help check for typical representations of missings. These are implemented in \texttt{naniar} as \texttt{common\_na\_numbers} and \texttt{common\_na\_strings}.

\hypertarget{verbs-replace-with}{%
\subsubsection{replace: Replacing values with missing values}\label{verbs-replace-with}}

Once possible missing values have been identified, these values can be replaced. For example, a dataset could have the values \texttt{-99} meaning a missing value. \texttt{naniar} implements replacement with the function, \texttt{replace\_with\_na()}. Values -99 could be replaced in the \texttt{x} column with:
\texttt{replace\_with\_na(dat\_ms,\ replace\ =\ list(x\ =\ -99))}. For operating on multiple variables, there are scoped variants for \texttt{replace\_with\_na}: \texttt{\_all}, \texttt{\_if}, and \texttt{\_at}. This means \texttt{replace\_with\_na\_all} operates on \textbf{all} columns, \texttt{replace\_with\_na\_at} operates \textbf{at} specific columns, and \texttt{replace\_with\_na\_if} makes a conditional change on columns \textbf{if} they meet some condition (such as \texttt{is.numeric} or \texttt{is.character}).

\hypertarget{verbs-add-cols}{%
\subsubsection{add: Adding missingness summary variables}\label{verbs-add-cols}}

Understanding the missingness structure can be improved by adding summary information alongside the data. For example, in \citet{Tierney2015}, the proportion of
missings in a row is used as the outcome in a model to identify variables
important in predicting missingness structures. \texttt{naniar} implements a series of functions to add these missingness summaries to the data, starting with \texttt{add\_}. These are inspired by the \texttt{add\_count} function in \texttt{dplyr}, which adds count information for specified groups or conditions. \texttt{naniar} provides operations to add the number or proportion of missingness, the missingness cluster, or the presence of any missings, with: \texttt{add\_n\_miss()}, \texttt{add\_prop\_miss()}, \texttt{add\_miss\_cluster}, and \texttt{add\_any\_miss()}, respectively (Table \ref{tab:add-missing-info}). There are also functions for adding information about shadow values, and readable labels for any missing values with \texttt{add\_label\_shadow()} and \texttt{add\_label\_missings()}.

\begin{CodeChunk}
\begin{table}

\caption{\label{tab:add-missing-info}Overview of the 'add' functions in naniar}
\centering
\begin{tabular}[t]{l|l}
\hline
Function & Adds column which:\\
\hline
add\_n\_miss(data) & Contains the number missing values in a row\\
\hline
add\_prop\_miss(data) & Contains the proportion of missing values in a row\\
\hline
add\_miss\_cluster(data) & Contains the missing value cluster\\
\hline
\end{tabular}
\end{table}

\end{CodeChunk}

\hypertarget{verbs-nabular}{%
\subsubsection{shadow: Creating nabular data}\label{verbs-nabular}}

\emph{Nabular} data has the shadow matrix column-bound to existing data. This facilitates visualization and summaries, and allows for imputed values to be tracked. \emph{Nabular} data can be created with \texttt{nabular()}:

\begin{CodeChunk}

\begin{CodeInput}
R> nabular(dat_ms)
\end{CodeInput}

\begin{CodeOutput}
# A tibble: 5 x 6
      x y         z x_NA  y_NA  z_NA 
  <dbl> <chr> <dbl> <fct> <fct> <fct>
1     1 A      -100 !NA   !NA   !NA  
2     3 N/A     -99 !NA   !NA   !NA  
3    NA <NA>    -98 NA    NA    !NA  
4   -99 E      -101 !NA   !NA   !NA  
5   -98 F        -1 !NA   !NA   !NA  
\end{CodeOutput}
\end{CodeChunk}

\hypertarget{verbs-recode}{%
\subsubsection{flag: Describing different types of missing values}\label{verbs-recode}}

Unusual or spurious data values are often identified and \texttt{flagged}. For example, there might be special codes to mark an individual dropping out of a study, known instrument failure in weather instruments, or for values censored in analysis. \texttt{naniar} provides tools to encode these special types of missingness in the shadow matrix of \emph{nabular} data, using \texttt{recode\_shadow()}. This requires specifying the variable to contain the flagged value, the condition for flagging, and a suffix. This is then recoded as a new factor level in the shadow matrix, so every column is aware of all possible new values of missingness. For example, -99 could be recoded to indicate a broken machine sensor for the variable \texttt{x} with the following:

\begin{CodeChunk}

\begin{CodeInput}
R> nabular(dat_ms) %>%
R+   recode_shadow(x = .where(x == -99 ~ "broken_sensor"))
\end{CodeInput}

\begin{CodeOutput}
# A tibble: 5 x 6
      x y         z x_NA             y_NA  z_NA 
  <dbl> <chr> <dbl> <fct>            <fct> <fct>
1     1 A      -100 !NA              !NA   !NA  
2     3 N/A     -99 !NA              !NA   !NA  
3    NA <NA>    -98 NA               NA    !NA  
4   -99 E      -101 NA_broken_sensor !NA   !NA  
5   -98 F        -1 !NA              !NA   !NA  
\end{CodeOutput}
\end{CodeChunk}

\hypertarget{verbs-impute}{%
\subsubsection{impute: Imputing values}\label{verbs-impute}}

\texttt{naniar} does not reinvent the wheel for imputation, instead working with existing methods. However, \texttt{naniar} provides a few imputation methods to facilitate exploration and visualization: \texttt{impute\_below}, \texttt{impute\_mean}, and \texttt{impute\_median}. While useful to explore structure in missingness, they are not recommended for use in analysis. \texttt{impute\_below} imputes values below the minimum value, with some controllable jitter (random noise) to reduce overplotting.

Similar to \texttt{simputation}, each \texttt{impute\_} function returns the data with values imputed. However, \texttt{naniar} does not use a formula syntax, instead each function has ``scoped variants'' \texttt{\_all}, \texttt{\_at} and \texttt{\_if} as in \ref{verbs-replace-with}. \texttt{impute\_} functions with no scoped variant, (\texttt{impute\_mean}), will work on a single vector, but not a data.frame. One challenge with this approach is imputed value locations are not tracked. This issue is resolved with \code{nabular} data covered in Section \ref{verbs-track}.

\hypertarget{verbs-track}{%
\subsubsection{track: Shadow and impute missing values}\label{verbs-track}}

To evaluate imputations they need to be tracked. This is achieved by first using \texttt{nabular} data, then imputing, and imputed values can then be referred to by their shadow variable, \texttt{\_NA} (Figure \ref{fig:track-impute-example}). The code below shows the track pattern, first using \texttt{nabular}, then imputing with \texttt{impute\_lm}. \texttt{label\_shadow} then adds a label to facilitate identifying missings:

\begin{CodeChunk}

\begin{CodeInput}
R> aq_imputed <- nabular(airquality) %>%
R+   as.data.frame() %>% 
R+   simputation::impute_lm(Ozone ~ Temp + Wind) %>%
R+   simputation::impute_lm(Solar.R ~ Temp + Wind) %>%
R+   add_label_shadow()
R> 
R> head(aq_imputed)
\end{CodeInput}

\begin{CodeOutput}
      Ozone  Solar.R Wind Temp Month Day Ozone_NA Solar.R_NA Wind_NA Temp_NA
1  41.00000 190.0000  7.4   67     5   1      !NA        !NA     !NA     !NA
2  36.00000 118.0000  8.0   72     5   2      !NA        !NA     !NA     !NA
3  12.00000 149.0000 12.6   74     5   3      !NA        !NA     !NA     !NA
4  18.00000 313.0000 11.5   62     5   4      !NA        !NA     !NA     !NA
5 -11.67673 127.4317 14.3   56     5   5       NA         NA     !NA     !NA
6  28.00000 159.5042 14.9   66     5   6      !NA         NA     !NA     !NA
  Month_NA Day_NA any_missing
1      !NA    !NA Not Missing
2      !NA    !NA Not Missing
3      !NA    !NA Not Missing
4      !NA    !NA Not Missing
5      !NA    !NA     Missing
6      !NA    !NA     Missing
\end{CodeOutput}
\end{CodeChunk}

Multiple missing or imputed values can be mapped to a graphical element in ggplot2 by setting the \texttt{color} or \texttt{fill} aesthetic in ggplot to \texttt{any\_missing}, a result of the \texttt{add\_label\_shadow()} function. (Figure \ref{fig:track-impute-example}). Imputed values can also be compared to complete case data, grouping by \texttt{any\_missing}, and then summarizing, similar to other dplyr summary workflows shown below in Table \ref{tab:impute-summary-out}, showing similarities and differences in imputation methods.

\begin{CodeChunk}
\begin{figure}

{\centering \includegraphics[width=1\linewidth]{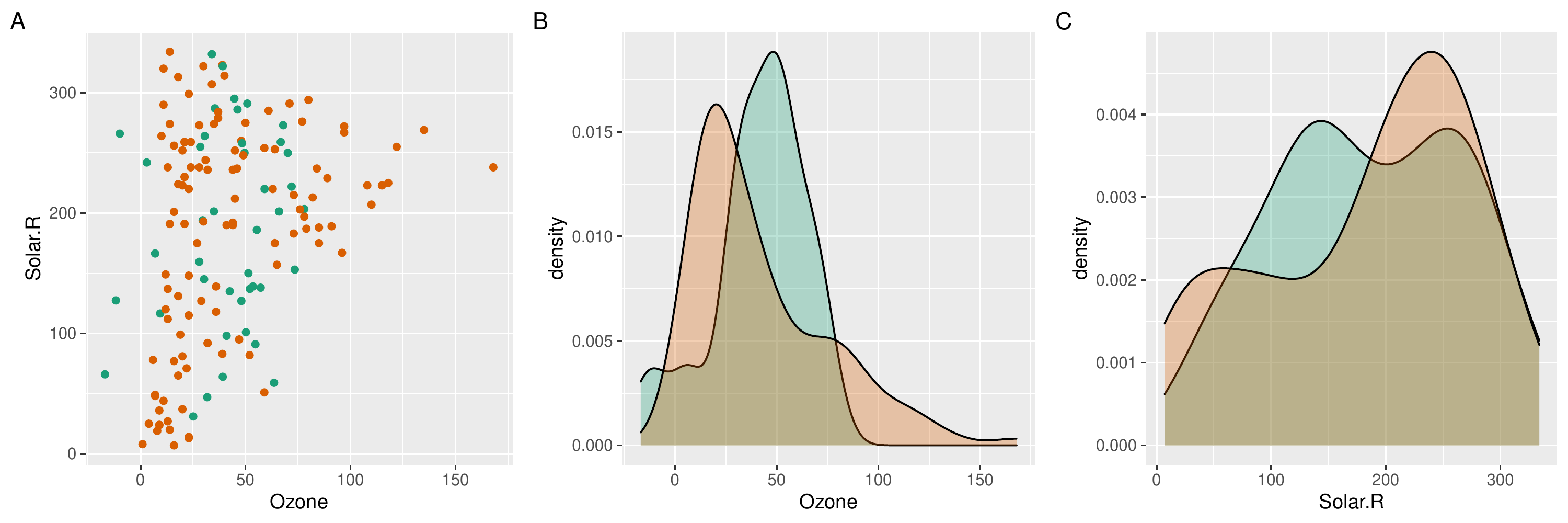} 

}

\caption[Scatterplot (A) and density plots (B and C) of ozone and solar radiation from the airquality dataset containing imputed values from a linear model]{Scatterplot (A) and density plots (B and C) of ozone and solar radiation from the airquality dataset containing imputed values from a linear model. Imputed values are colored green, and data values orange. Imputed values are similar, but slightly trended to the mean.}\label{fig:track-impute-example}
\end{figure}
\end{CodeChunk}

\begin{CodeChunk}

\begin{CodeInput}
R> aq_imputed %>%
R>   group_by(any_missing) %>%
R>   summarise_at(.vars = vars(Ozone),
R>                .funs = lst(min, mean, median, max)) 
\end{CodeInput}
\end{CodeChunk}

\begin{CodeChunk}
\begin{table}

\caption{\label{tab:impute-summary-out}Summarising values of imputed vs non imputed values. Comparing imputed values (denoted as 'Missing', since they were previously missing), the mean and median values are similar, but the minimum and maximum values are very different.}
\centering
\begin{tabular}[t]{l|r|r|r|r}
\hline
any\_missing & min & mean & median & max\\
\hline
Missing & -16.86418 & 41.22494 & 45.4734 & 78\\
\hline
Not Missing & 1.00000 & 42.09910 & 31.0000 & 168\\
\hline
\end{tabular}
\end{table}

\end{CodeChunk}

\hypertarget{graphics}{%
\section{Graphics}\label{graphics}}

Missing values are often ignored when plotting data - which is why the data visualization software, MANET, was named and is an acronym corresponding to ``Missings Are Now Equally Treated'' \citep{Unwin1996}. However, plots can help to identify the type of missing value patterns, even those of MCAR, MAR or MNAR. Here we summarise how to systematically explore missing patterns visually, and define useful plots to make, relative to the nabular data structure.

\hypertarget{overviews}{%
\subsection{Overviews}\label{overviews}}

The first step is to get an overview of the extent of the missingness. Overview visualizations for variables and cases are provided with \texttt{gg\_miss\_var} and \texttt{gg\_miss\_case} (Figure \ref{fig:gg-miss-case-var}A), drawing attention to the amount of missings, and ordering by missingness. The ``airquality'' dataset (from base R), is shown, and contains daily air quality measurements in New York, from May to September, 1973. We learn from figure \ref{fig:gg-miss-case-var}A-B, that two variables contain missings, approximately one third of observations have one missing value, and a tiny number of observations are missing across two variables. These overview plots are created from the shadow matrix in long form (Figure \ref{fig:nabularfig}). Numerical statistics can also be reported (Section \ref{num-sum}).

\begin{CodeChunk}
\begin{figure}

{\centering \includegraphics[width=0.9\linewidth]{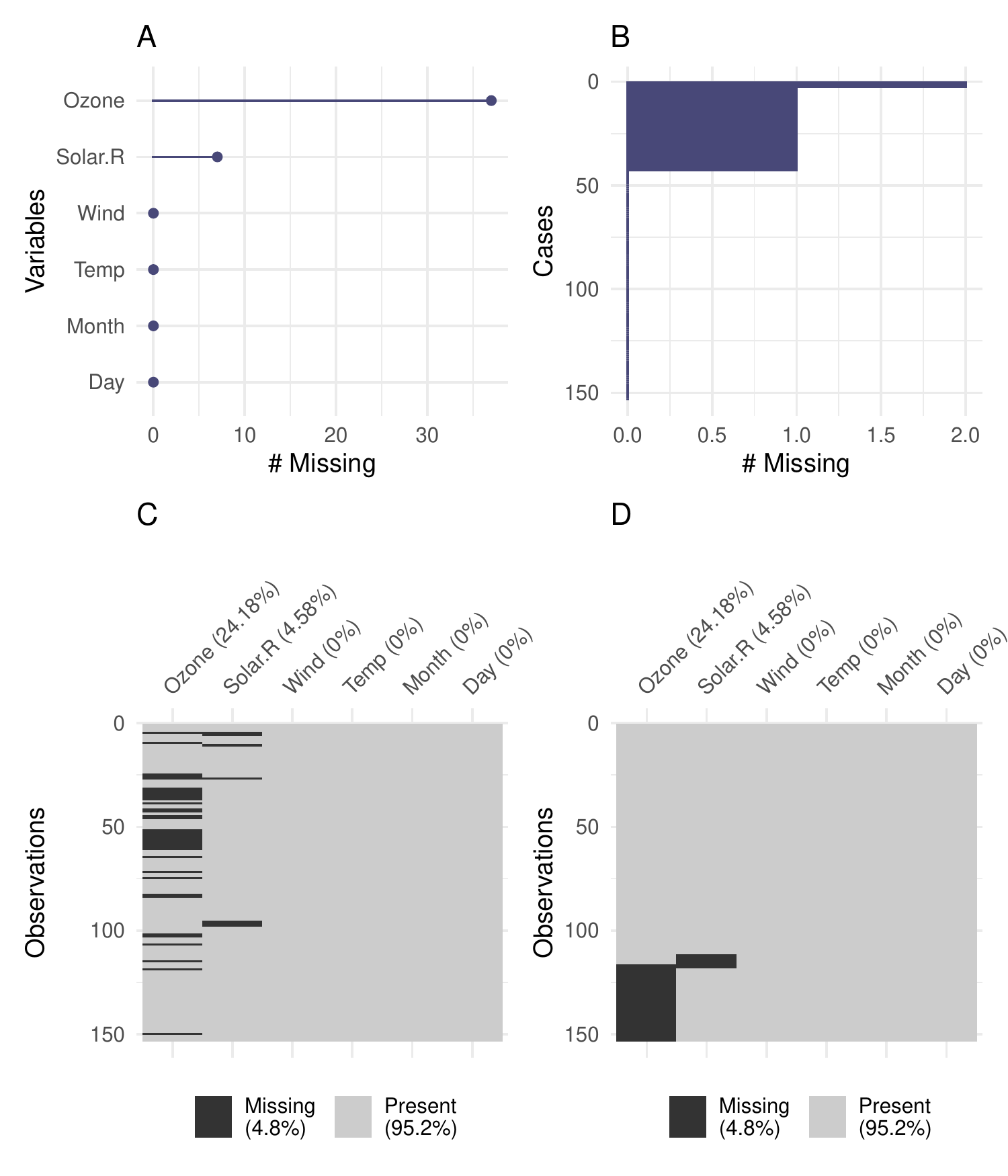} 

}

\caption[Graphical summaries of missingness in the airquality data]{Graphical summaries of missingness in the airquality data. Missings invariables (A) and cases (B), and for a birds eye view with missingness as a heatmap in (C), and with clustering applied (D). There are missing values in Ozone and Solar.R, with Ozone having more missings. Not many cases have two missings. Most missingness is from cases with one missing value. The default output (C) and ordered by clustering on rows and columns (D). These overviews are made possible using the shadow matrix in long form. There are only missings in ozone and solar radiation, and there appears to be some structure to their missingness.}\label{fig:gg-miss-case-var}
\end{figure}
\end{CodeChunk}

The shadow matrix can be put into long form, allowing both the variables and cases to be displayed using a heatmap style visualization, with \texttt{vis\_miss()} from \pkg{visdat} \citep{visdat} (Figure \ref{fig:nabularfig}). This also provides numerical summaries of missingness in the legend, and for each column (Figure \ref{fig:gg-miss-case-var}C-D). Clustering can be applied to the rows, and columns arranged in order of missingness (Figure \ref{fig:gg-miss-case-var}D). Similar visualizations are available in other packages such as \pkg{VIM}, \pkg{mi}, \pkg{Amelia}, and \pkg{MissingDataGUI}. A key improvement is \texttt{vis\_miss()} orients the visualization analogous to a regular data structure: variables form columns and are named at the top, and each row is an observation. Using \pkg{ggplot2}, as the foundation, makes the plot easily customizable.

\begin{CodeChunk}
\begin{figure}

{\centering \includegraphics{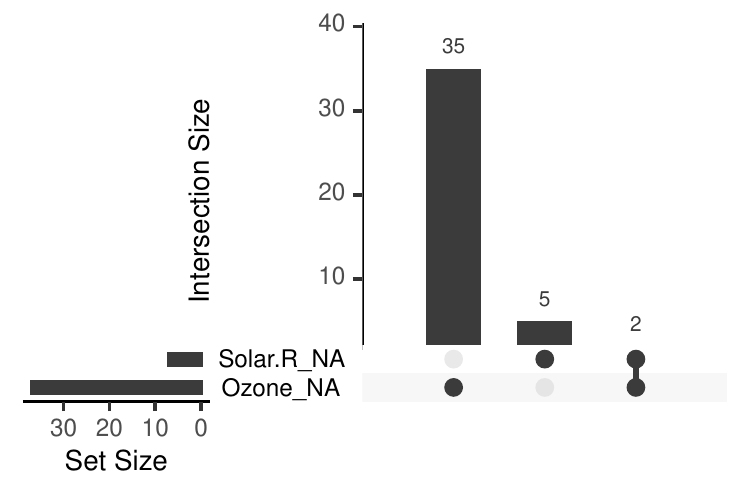} 

}

\caption[The pattern of missingness in the airquality dataset shown in an upset plot]{The pattern of missingness in the airquality dataset shown in an upset plot. Only Ozone and Solar.R have missing values, and Ozone has the most missing values. There are 2 cases where both Solar.R and Ozone have missing values.}\label{fig:airquality-upset}
\end{figure}
\end{CodeChunk}

The number of times observations are missing together can be visualized using an ``upset plot'' \citep{Conway2017}. An alternative to a Venn diagram, an upset plot shows the size and features of overlapping sets, and scales well with more variables. An upset plot can be constructed from the shadow matrix, as shown in Figure \ref{fig:airquality-upset} which displays the overlapping counts of missings in the airquality data. The bottom right shows the combinations of missingness, the top panel shows the size of these combinations, and the bottom left shows missingness in each variable. This provides similar information to Figure \ref{fig:gg-miss-case-var}A-D, but more clearly illustrating overlapping missingness, where 2 cases are missing together in variables Solar.R and Ozone.

\hypertarget{univariate}{%
\subsection{Univariate}\label{univariate}}

Missing values are typically not shown for univariate visualizations such as histograms or densities. Two ways to use \emph{nabular} data to present univariate data with missings are discussed. The first imputes values below the range to facilitate visualizations. The second displays two plots of the same variable according to the missingness of a chosen variable.

\textbf{Imputing values below the range}. To visualize the amount of missings in each variable, the data is transformed into \emph{nabular} form, then values are imputed below the range of data using \texttt{impute\_below\_all}. Figure \ref{fig:impute-shift-histogram}A shows a histogram of Ozone values on the right in green, and the histogram of missing ozone values on the left, in orange. The missings in Ozone are imputed and \texttt{Ozone\_NA} is mapped to the fill aesthetic. Nabular data facilitates adding counts of missingness to a histogram, allowing examination of a variables' distribution of values, and also the magnitude of missings.

\textbf{Univariate split by missingness}. The distribution of a variable can be shown according to the missingness of another variable. The shadow matrix part of nabular is used to handle the faceting, and color mapping. Figure \ref{fig:impute-shift-histogram} shows the values of temperature when ozone is present, and missing, using a faceted histogram (B), and an overlaid density (C). This shows how values of temperature are affected by the missingness of ozone, and reveals a cluster of low temperature observations with missing ozone values. This type of plot can facilitate exploring missing data distributions. For example, we would expect if data were MCAR, for values to be roughly uniformly missing throughout the histogram or density.

\begin{CodeChunk}
\begin{figure}

{\centering \includegraphics[width=0.75\linewidth]{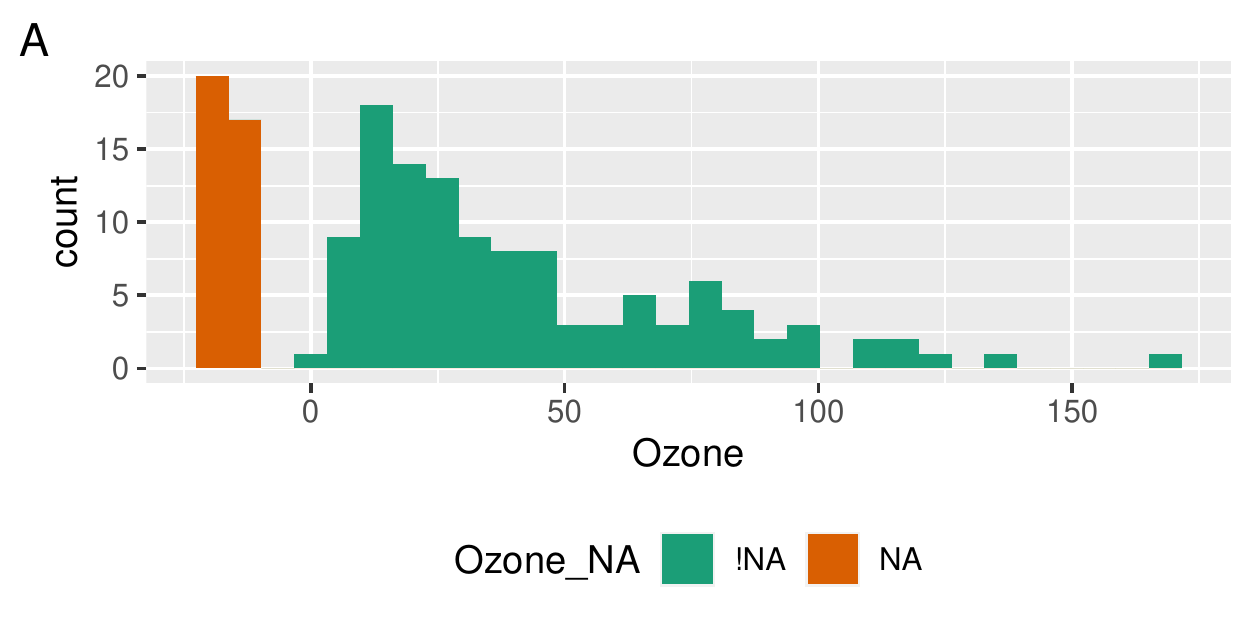} \includegraphics[width=0.75\linewidth]{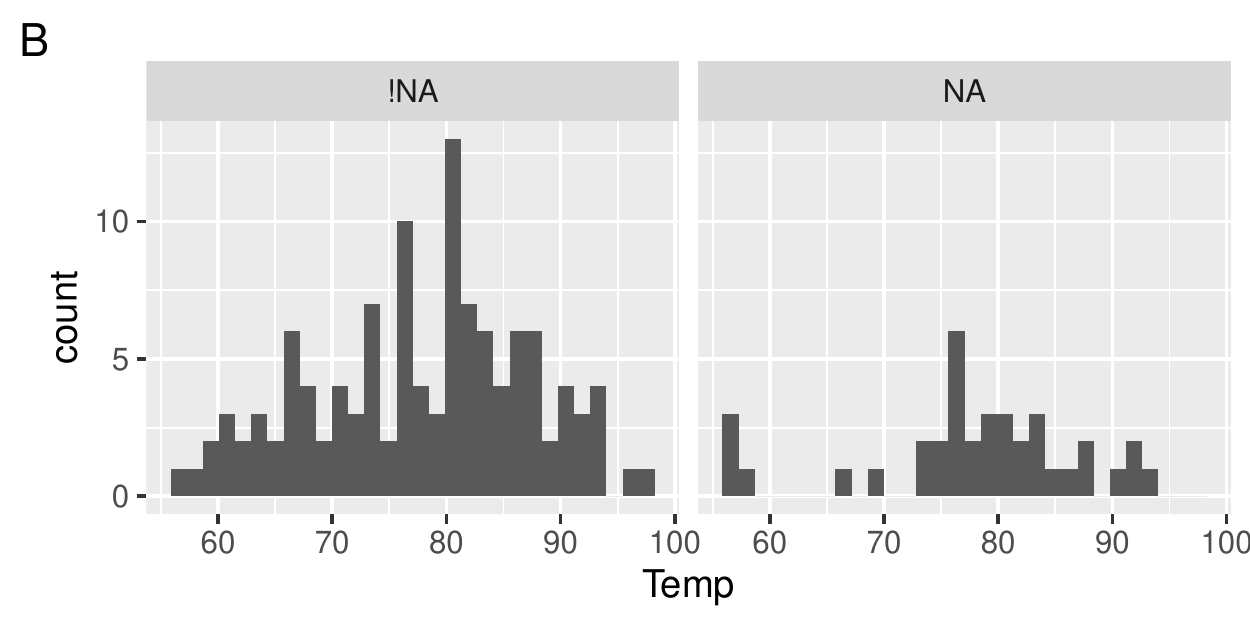} \includegraphics[width=0.75\linewidth]{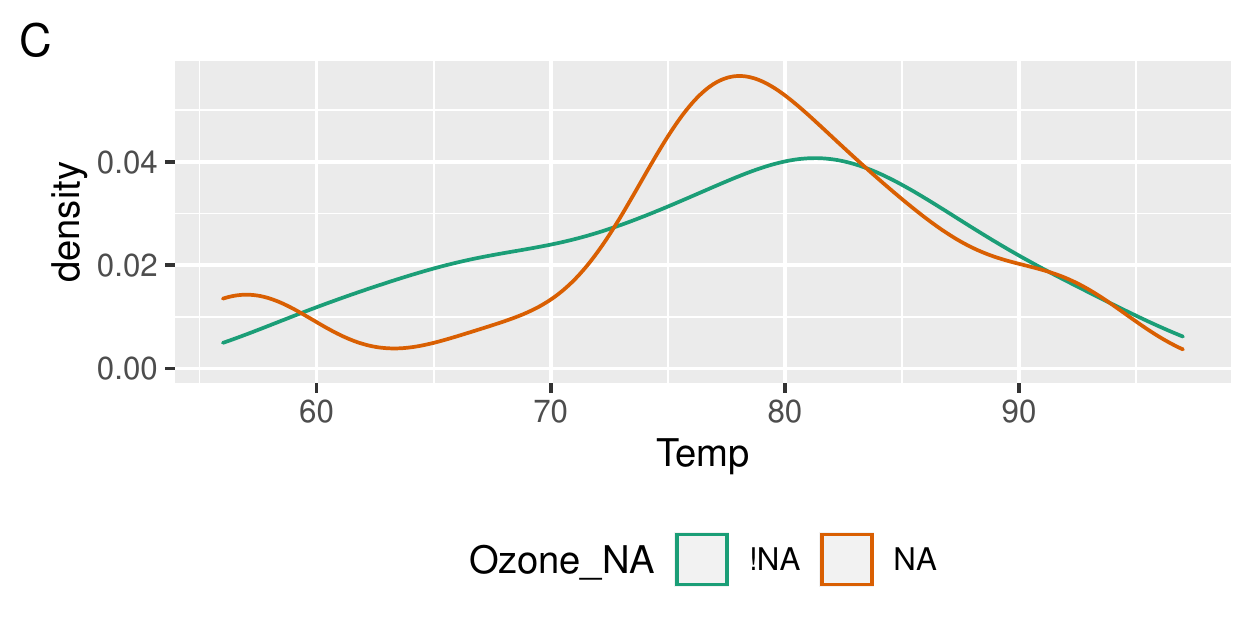} 

}

\caption[Univariate summaries of missingness]{Univariate summaries of missingness. (A) A histogram using nabular data to show the values and missings in ozone. Values are imputed below the range to show the number of missings in ozone and colored according to missingness of ozone (`Ozone\_NA`). There are about 35 missings in Ozone. Panel C shows temperature according to missingness in ozone from in the airquality dataset. A histogram of temperature facetted by the missingness of ozone (B), or a density of temperature colored by missingness in ozone (C). These show a cluster of low temperature observations with missing ozone values, but temperature is otherwise similar.}\label{fig:impute-shift-histogram}
\end{figure}
\end{CodeChunk}

\hypertarget{bivariate}{%
\subsection{Bivariate}\label{bivariate}}

To visualize missing values in two dimensions the missing values can be placed in plot margins, by imputing values below the range of the data. Using \emph{nabular} data identifies imputed values, and color makes missingness pre-attentive \citep{treisman1985}. The steps of imputing and coloring are combined into \texttt{geom\_miss\_point()}. Figure \ref{fig:geom-miss} shows a mostly uniform spread of missing values for Solar.R and Ozone. As \texttt{geom\_miss\_point()} is a defined \pkg{ggplot2} geometry, it works with features such as faceting and mapping other variables to graphical aesthetics.

\begin{CodeChunk}
\begin{figure}

{\centering \includegraphics{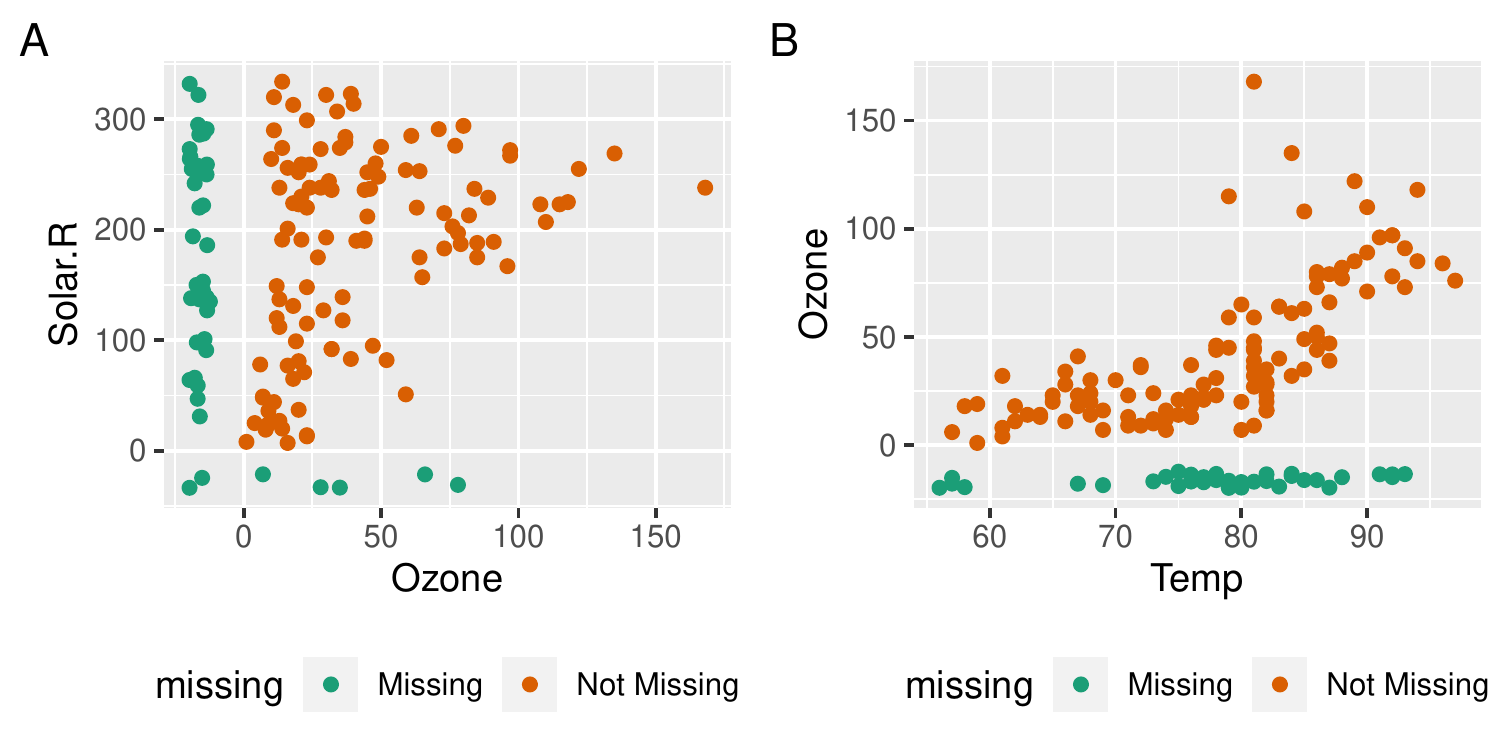} 

}

\caption[Scatterplots with missings displayed at 10 percent below for the airquality dataset]{Scatterplots with missings displayed at 10 percent below for the airquality dataset. Scatterplots of ozone and solar radiation (A), and ozone and temperature (B). There are missings in ozone and solar radiation, but not temperature.}\label{fig:geom-miss}
\end{figure}
\end{CodeChunk}

\hypertarget{multivariate}{%
\subsection{Multivariate}\label{multivariate}}

Parallel coordinate plots can help to visualize missingness beyond two dimensions. They transform variables to the same scale, ranging between 0 and 1. The \texttt{oceanbuoys} dataset from \pkg{naniar} is used for this visualization, containing measurements of sea and air temperature, humidity, and east west and north south wind speeds. Data was collected in 1993 and 1997, to understand and predict El Niño and El Niña. Figure \ref{fig:parallel-cord-plot} is a parallel coordinate plot of \texttt{oceanbuoys}, with missing values imputed to be 10\% below the range, and values colored according to whether humidity was missing (\texttt{humidity\_NA}). Figure \ref{fig:parallel-cord-plot} shows humidity is missing at low air and sea temperatures, and humidity is missing in one year, and one location.

\begin{CodeChunk}
\begin{figure}

{\centering \includegraphics[width=1\linewidth]{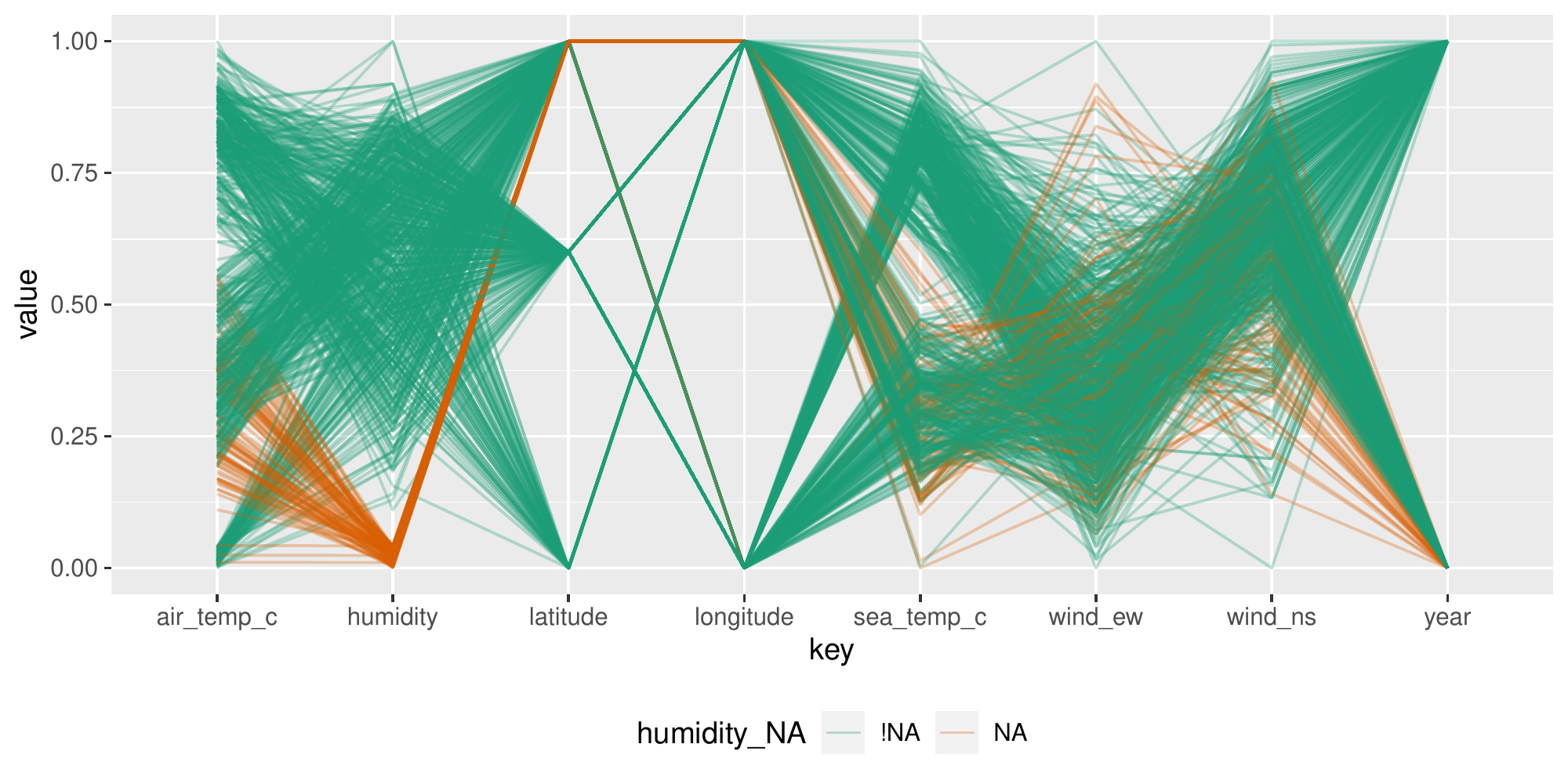} 

}

\caption[Parallel coordinate plot shows missing values imputed 10\% below range for the oceanbuoys dataset]{Parallel coordinate plot shows missing values imputed 10\% below range for the oceanbuoys dataset. Values are colored by missingness of humidity. Humidity is missing for low air and sea temperatures, and is missing for one year and one location. }\label{fig:parallel-cord-plot}
\end{figure}
\end{CodeChunk}

\hypertarget{num-sum}{%
\section{Numerical summaries}\label{num-sum}}

This section describes approaches to summarizing missingness, and an implementation in \pkg{naniar}. Numerical summaries should be easy to remember with consistent names and output, returning either a single number \ref{single-num-sum}, or a dataframe \ref{sum-tab-missings}, so they integrate well with plotting and modelling tools. How these work with other tools in an analysis pipeline is shown in \ref{num-sum-w-group}.

\hypertarget{single-num-sum}{%
\subsection{Single number summaries}\label{single-num-sum}}

The overall number, proportion, or percent of missing values in a dataset should be simple to calculate. \texttt{naniar} provides the functions \texttt{n\_miss}, \texttt{prop\_miss} and \texttt{pct\_miss}, as well as their complements. Summaries for variables and cases are made by appending \texttt{\_case} or \texttt{\_var} to these summaries. An overview is shown in Table \ref{tab:n-prop-pct-miss-complete}.

\begin{CodeChunk}
\begin{table}

\caption{\label{tab:n-prop-pct-miss-complete}Single number summaries of missingness and completeness of the airquality dataset. The functions follow consistent naming, making them easy to remember, and their use clear.}
\centering
\begin{tabular}[t]{l|r|l|r}
\hline
Missing Function & missing value & Complete function & complete value\\
\hline
n\_miss & 44.00 & n\_complete & 874.00\\
\hline
prop\_miss & 0.05 & prop\_complete & 0.95\\
\hline
pct\_miss & 4.79 & pct\_complete & 95.21\\
\hline
pct\_miss\_case & 27.45 & prop\_complete\_case & 72.55\\
\hline
pct\_miss\_var & 33.33 & pct\_complete\_var & 66.67\\
\hline
\end{tabular}
\end{table}

\end{CodeChunk}

\hypertarget{sum-tab-missings}{%
\subsection{Summaries and tabulations of missing data}\label{sum-tab-missings}}

Presenting the number and percent of missing values for each variable, or case, provides a summary usable in data handling, or in models to inform imputation. For example, potentially dropping variables, or deciding to include others in an imputation model. Another useful approach is to tabulate the frequency of missing values for each variable or case; that is, the number of times there are zero missings, one missing, two, and so on. These summaries and tabulations are shown for variables in Tables \ref{tab:miss-var-summary} and \ref{tab:miss-var-table}, and implemented with \texttt{miss\_var\_summary} and \texttt{miss\_var\_table}. Case-wise (row-wise) summaries and tabulations are implemented with \texttt{miss\_case\_summary} and \texttt{miss\_case\_table}.

These summaries order rows by the number of missings (\texttt{n\_miss}), to show the most missings at the top. The number of missings across a repeating span, or finding ``runs'' or ``streaks'' of missings in given variables can also be useful for identifying missingness patterns, and are implemented with \texttt{miss\_var\_span}, and \texttt{miss\_var\_run}.

\begin{CodeChunk}
\begin{table}

\caption{\label{tab:miss-var-summary}\texttt{miss\char`_var\char`_summary} provides the number and percent of missings in each variable in airquality. Only ozone and solar radiation have missing values.}
\centering
\begin{tabular}[t]{l|r|r}
\hline
variable & n\_miss & pct\_miss\\
\hline
Ozone & 37 & 24.2\\
\hline
Solar.R & 7 & 4.6\\
\hline
Wind & 0 & 0.0\\
\hline
Temp & 0 & 0.0\\
\hline
Month & 0 & 0.0\\
\hline
Day & 0 & 0.0\\
\hline
\end{tabular}
\end{table}

\end{CodeChunk}

\begin{CodeChunk}
\begin{table}

\caption{\label{tab:miss-var-table}\texttt{miss\char`_var\char`_table} tabulates the amount of missing data in each variable in airquality. This shows the number of variables with 0, 7, and 37 missings, and the percentage of variables with those amounts of missingness. There are few missingness patterns.}
\centering
\begin{tabular}[t]{r|r|r}
\hline
n\_miss\_in\_var & n\_vars & pct\_vars\\
\hline
0 & 4 & 66.7\\
\hline
7 & 1 & 16.7\\
\hline
37 & 1 & 16.7\\
\hline
\end{tabular}
\end{table}

\end{CodeChunk}

\hypertarget{num-sum-w-group}{%
\subsection{Combining numerical summaries with grouping operations}\label{num-sum-w-group}}

It is useful to explore summaries and tabulations within groups of a dataset. \texttt{naniar} works with \texttt{dplyr}'s \texttt{group\_by} operator to produce grouped summaries, which work well with the ``pipe'' operator. The code and table below \ref{tab:group-miss-var-summary} show an example of missing data summaries for airquality, grouped by month.

\begin{CodeChunk}
\begin{table}

\caption{\label{tab:group-miss-var-summary}\texttt{miss\char`_var\char`_summary} combined with \texttt{group\char`_by} provides a grouped summary of the missingness in each variable, for each Month of the airquality dataset. Only the first 10 rows are shown. There are more ozone missings in June than May.}
\centering
\begin{tabular}[t]{r|l|r|r}
\hline
Month & variable & n\_miss & pct\_miss\\
\hline
5 & Ozone & 5 & 16.1\\
\hline
5 & Solar.R & 4 & 12.9\\
\hline
5 & Wind & 0 & 0.0\\
\hline
5 & Temp & 0 & 0.0\\
\hline
5 & Day & 0 & 0.0\\
\hline
6 & Ozone & 21 & 70.0\\
\hline
6 & Solar.R & 0 & 0.0\\
\hline
6 & Wind & 0 & 0.0\\
\hline
6 & Temp & 0 & 0.0\\
\hline
6 & Day & 0 & 0.0\\
\hline
\end{tabular}
\end{table}

\end{CodeChunk}

\hypertarget{case-study}{%
\section{Application}\label{case-study}}

This section shows how the methods described so far are used together in a data analysis workflow. We analyse a case study of housing data for the city of Melbourne from January 28, 2016 to March 17, 2018. The data was compiled by scraping weekly property clearance data \citep{Kaggle-2018-data}. There are 27,247 properties, and 21 variables in the dataset. The variables include real estate type (town house, unit, house), suburb, selling method, number of rooms, price, real estate agent, sale date, and distance from the Central Business District (CBD).

The goal in analyzing this data is to accurately predict Melbourne housing prices. The data contains many missing values. As a precursor to building a predictive model, this analysis focuses on understanding the patterns of missingness.

\hypertarget{case-study-explore-pattern}{%
\subsection{Exploring patterns of missingness}\label{case-study-explore-pattern}}

Figure \ref{fig:housing-miss-case-var}A shows 9 variables with missing values. The most missings are in ``building area'', followed by ``year built'', and ``land size'', with similar amounts of missingness in ``Car'', ``bathroom'', ``bedroom2'', ``longitude'', and ``latitude.'' Figure \ref{fig:housing-miss-case-var}B reveals there are up to 50\% missing values in cases, and the majority of cases have more than 5\% values missing. The variables ``building area'' and ``year built'' have more than 50\% missing data, and so could perhaps be omitted from subsequent analysis, as imputed values are likely to be spurious. Three missingness clusters are revealed by visualizing missingness in the whole dataset, clustering and arranging the rows and columns of the data (Figure \ref{fig:applic-vis-miss}).

\begin{CodeChunk}
\begin{figure}

{\centering \includegraphics[width=1\linewidth]{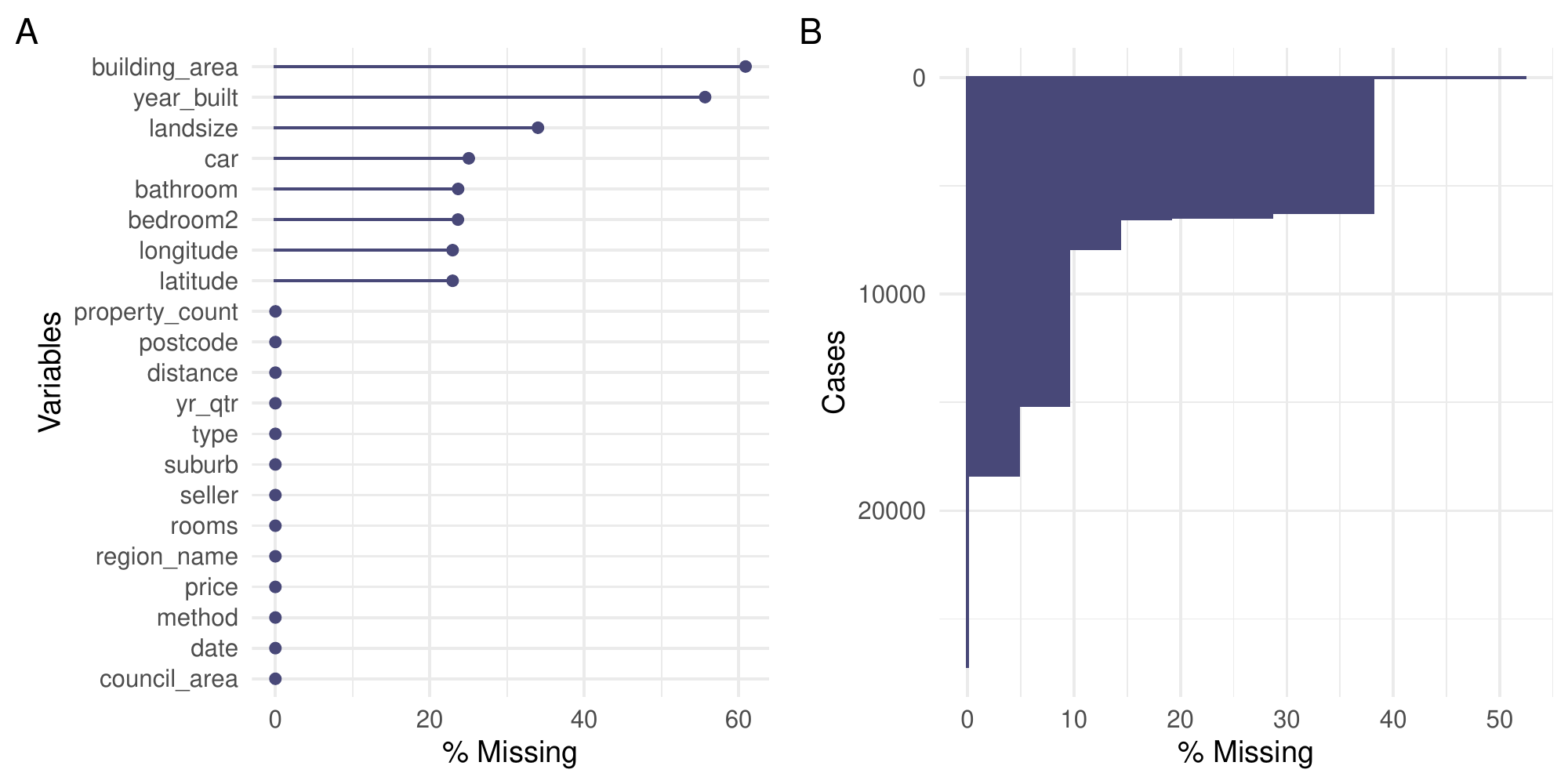} 

}

\caption[The amount of missings in variables (A) and cases (B) for Melbourne housing data]{The amount of missings in variables (A) and cases (B) for Melbourne housing data. (A) Build area and year built have more than 50\% missing, and car, bathroom, bedroom2 and longitude and latitude have about 25\% missings. (B) Cases are missing 5 - 50\% of values. The majority of missingness is in selected cases and variables.}\label{fig:housing-miss-case-var}
\end{figure}
\end{CodeChunk}

\begin{CodeChunk}
\begin{figure}

{\centering \includegraphics[width=0.85\linewidth]{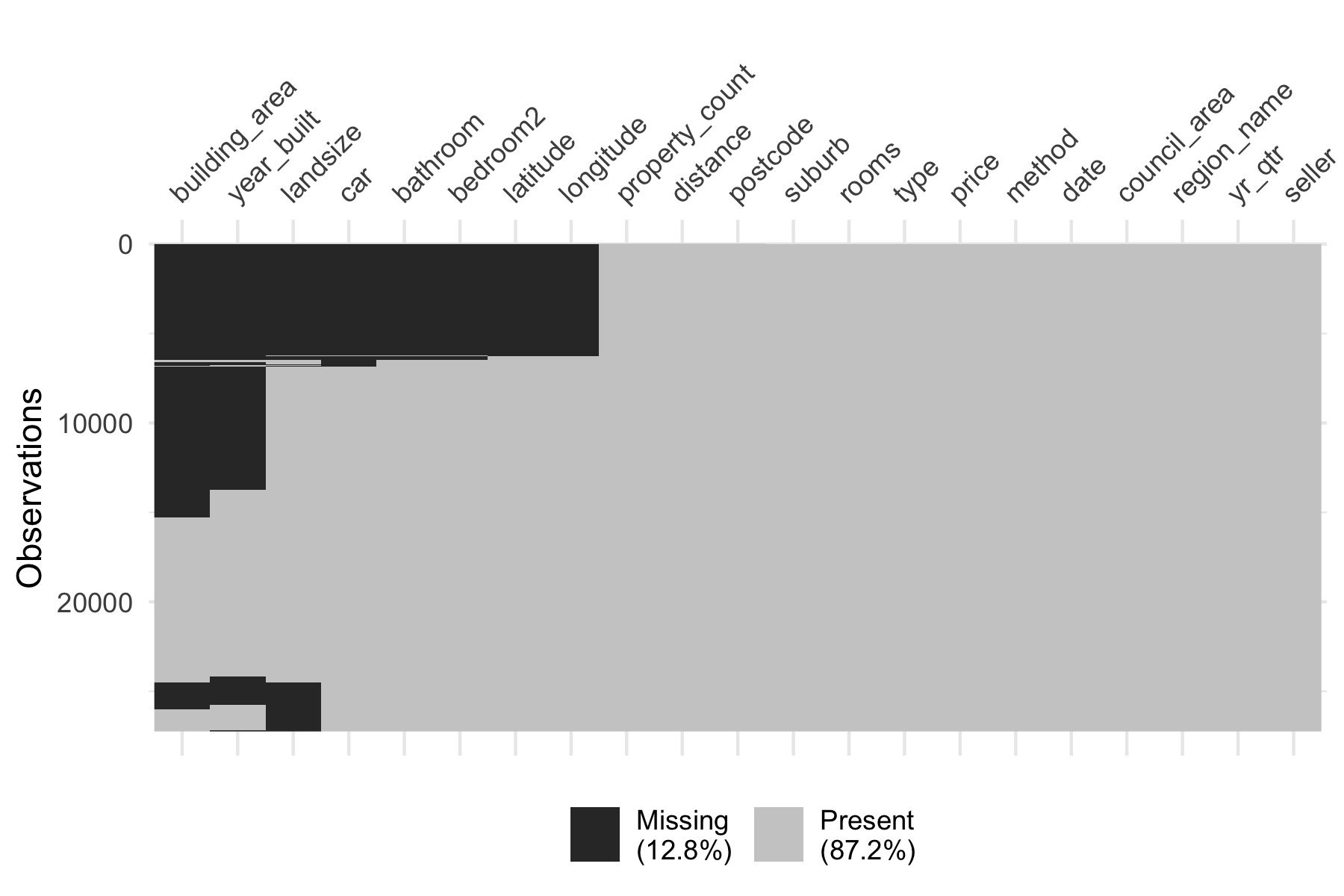} 

}

\caption[Heatmap of clustered missingness for housing data reveals structured missingness]{Heatmap of clustered missingness for housing data reveals structured missingness. Three groups of missingness are apparent. At the top: building area to longitude; the middle: building area and year built; the end: building area, year built, and landsize.}\label{fig:applic-vis-miss}
\end{figure}
\end{CodeChunk}

Figure \ref{fig:housing-upset} shows missingness patterns with an \texttt{upset} plot \citep{Conway2017}, displaying 8 intersecting sets of missing variables. Two patterns stand out: two, and five variables missing, providing further evidence of the missingness patterns seen in Figures \ref{fig:housing-miss-case-var} and \ref{fig:housing-upset}.

\begin{CodeChunk}
\begin{figure}

{\centering \includegraphics[width=1\linewidth]{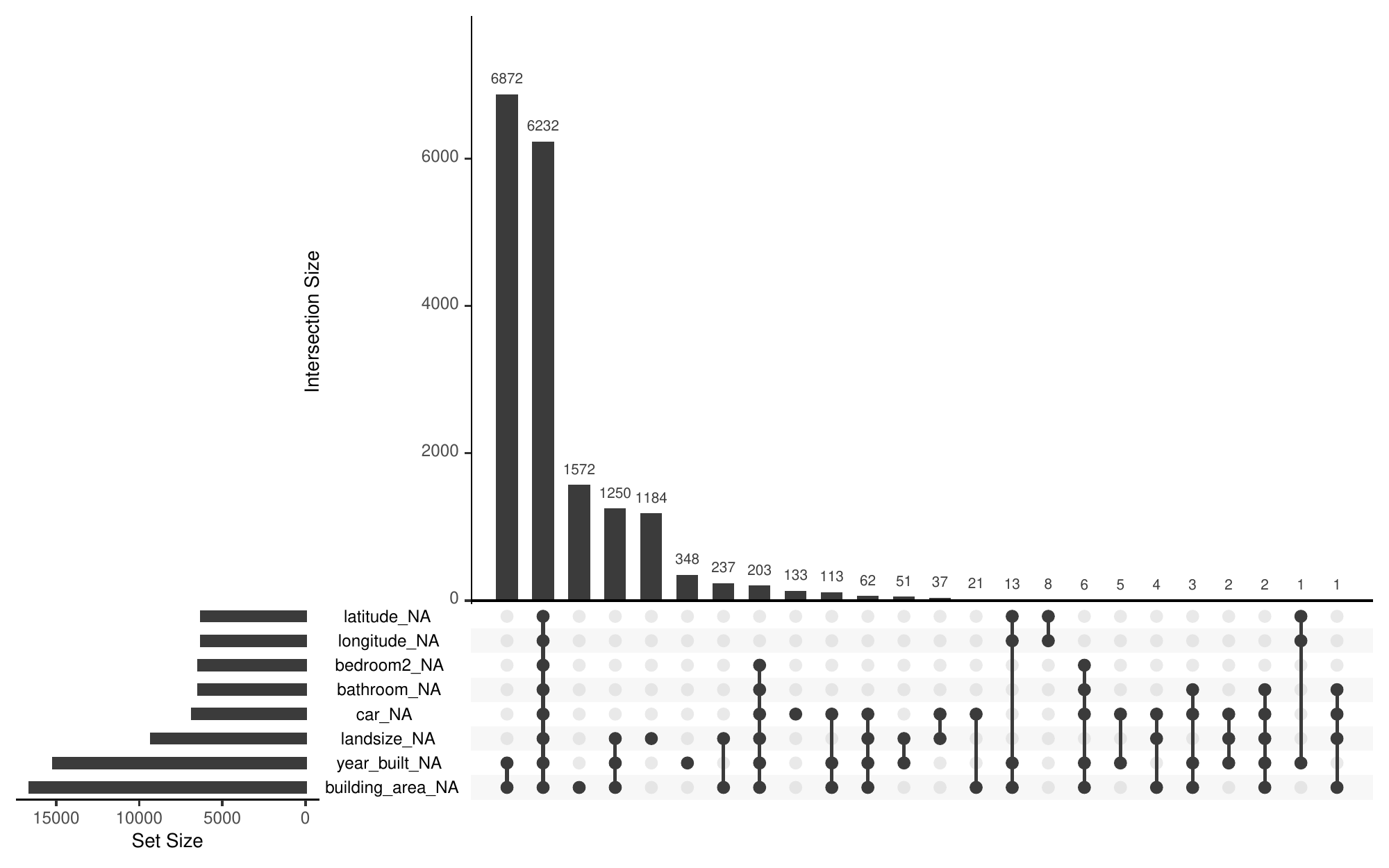} 

}

\caption[An upset plot of 8 sets of missingness in the housing data]{An upset plot of 8 sets of missingness in the housing data. Missingness for each variable is shown on the bottom left. Connected dots show co-occurences of missings in variables. Two missingness patterns are clear, year built and building area, and lattitude through to building area.}\label{fig:housing-upset}
\end{figure}
\end{CodeChunk}

Tabulating the number of missings in variables in Table \ref{tab:housing-miss-var-case-table} (left) shows three groups of missingness. Tabulating missings in cases (Table \ref{tab:housing-miss-var-case-table} (right)) shows six patterns of missingness. These overview plots lead to the removal of two variables from with more than 50\% missingness from analysis: ``building area'' and ``year built''.

\begin{CodeChunk}
\begin{table}
\caption{\label{tab:housing-miss-var-case-table}Tabulating missingness for variables (left) and cases (right) to understand missingness patterns. 14 variables have 0 - 3 missings, 6 variables have 6000 - 9000 missings, and 2 variables have 15 - 16,000 missings. About 30\% of cases have no missings, 45\% of cases have 1 - 6 missings, and about 23\% of cases have 8 or more missings. There are different patterns of missingness in variables and cases, but they can be broken down into smaller groups.}

\centering
\begin{tabular}[t]{r|r|r}
\hline
n\_miss\_in\_var & n\_vars & pct\_vars\\
\hline
0 & 10 & 47.6\\
\hline
1 & 2 & 9.5\\
\hline
3 & 1 & 4.8\\
\hline
6254 & 2 & 9.5\\
\hline
6441 & 1 & 4.8\\
\hline
6447 & 1 & 4.8\\
\hline
6824 & 1 & 4.8\\
\hline
9265 & 1 & 4.8\\
\hline
15163 & 1 & 4.8\\
\hline
16591 & 1 & 4.8\\
\hline
\end{tabular}
\centering
\begin{tabular}[t]{r|r|r}
\hline
n\_miss\_in\_case & n\_cases & pct\_cases\\
\hline
0 & 8887 & 32.6\\
\hline
1 & 3237 & 11.9\\
\hline
2 & 7231 & 26.5\\
\hline
3 & 1370 & 5.0\\
\hline
4 & 79 & 0.3\\
\hline
5 & 8 & 0.0\\
\hline
6 & 203 & 0.7\\
\hline
8 & 6229 & 22.9\\
\hline
9 & 2 & 0.0\\
\hline
11 & 1 & 0.0\\
\hline
\end{tabular}
\end{table}

\end{CodeChunk}

\hypertarget{case-study-explore-for-imp}{%
\subsection{Exploring missingness patterns for imputation}\label{case-study-explore-for-imp}}

Using information from \ref{case-study-explore-pattern}, the following variables are explored for features predicting missingness: ``Land size'', ``latitude'', ``longitude'', ``bedroom2'', ``bathroom'', ``car'', and ``land size''.

Missingness structure is explored by clustering the missing values into groups. Then, a classification and regression trees (CART) model is applied to predict these missingness clusters using the remaining variables \citep{Tierney2015, Barnett2017}. Two clusters of missingness are identified and predicted using all variables in the dataset with the CART package \texttt{rpart} \citep{rpart}, and plotted using the \texttt{rpart.plot} package \citep{rpart-plot}. Importance scores reveal the following variables as most important in predicting missingness: ``rooms'', ``price'', ``suburb'', ``council area'', ``distance'', and ``region name''. These variables are important for predicting missingness, so are included in the imputation model.

\begin{CodeChunk}
\begin{figure}

{\centering \includegraphics[width=0.9\linewidth]{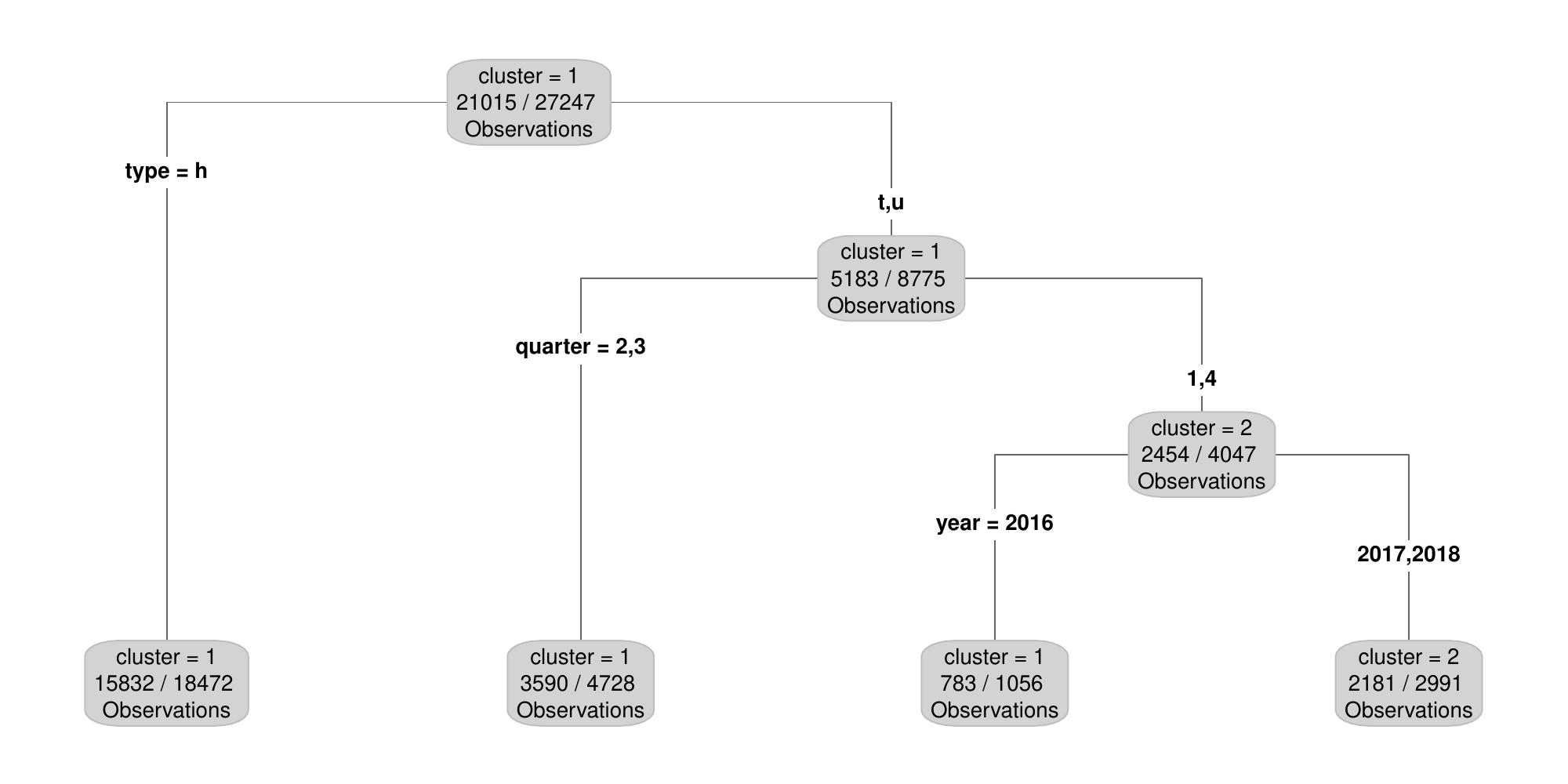} 

}

\caption[Decision tree output predicting missingness clusters]{Decision tree output predicting missingness clusters. Type of house, year quarter, and year were important for predicting missingness cluster. The cluster with the most missingness was for quarters 1 and 4, for 2017 and 2018. Type of house, year, and year quarter are important features related to missingness structure.}\label{fig:rpart-plot}
\end{figure}
\end{CodeChunk}

\hypertarget{case-study-imp-diagnosis}{%
\subsection{Imputation and diagnostics}\label{case-study-imp-diagnosis}}

\pkg{simputation} is used to implement two imputations: simple linear regression and K nearest neighbors. Values are imputed stepwise in ascending order of missingness. The track missings pattern is applied (described in \ref{verbs-track}), to assess imputed values. Imputed datasets are compared on their performance in a model predicting log house price for 4 variables (Figure \ref{fig:imputed-by-model}). Compared to KNN imputed values, the linear model imputed values closer to the mean.

\begin{CodeChunk}
\begin{figure}

{\centering \includegraphics[width=0.95\linewidth]{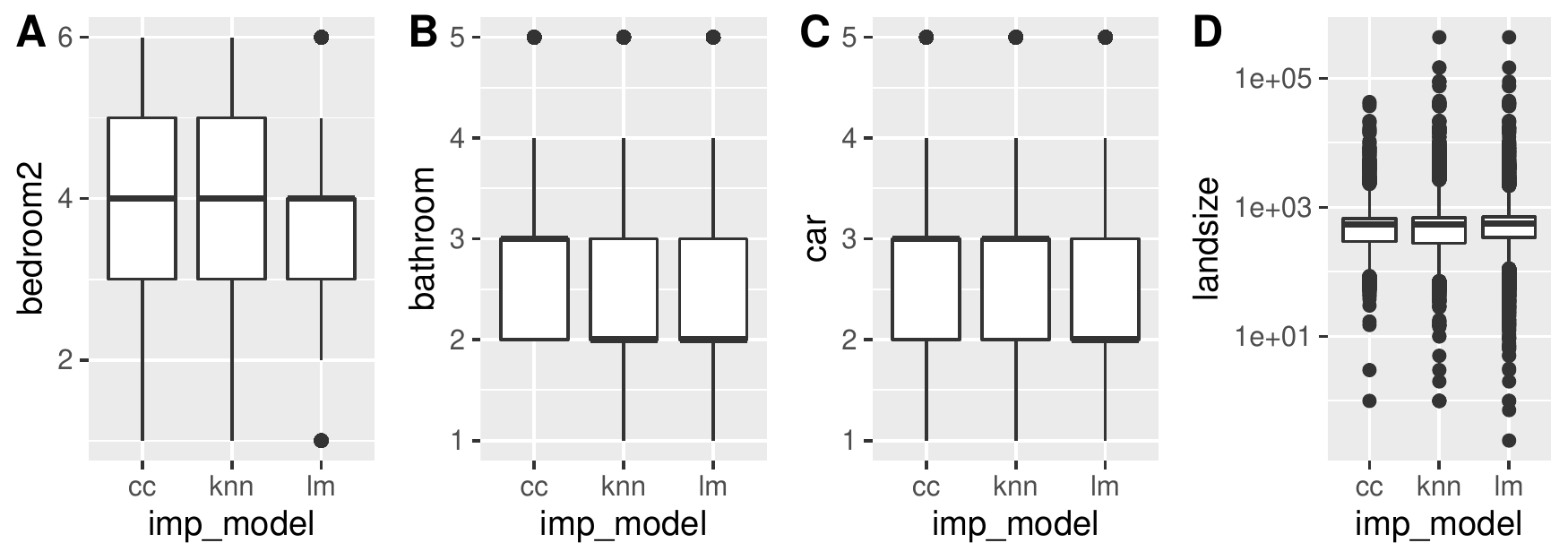} 

}

\caption[Boxplots of complete case data, and data imputed with KNN or linear model for different variables]{Boxplots of complete case data, and data imputed with KNN or linear model for different variables. (A) number of bedrooms, (B) number of bathrooms, (C) number of carspots, and (D) landsize (on a log10 scale). KNN had similar results to complete case, and linear model had a lower median for cars and fewer extreme values for bedrooms.}\label{fig:imputed-by-model}
\end{figure}
\end{CodeChunk}

\hypertarget{case-study-assess-model}{%
\subsection{Assessing model predictions}\label{case-study-assess-model}}

Coefficients of the linear model of log price vary for room for different imputed datasets (Figure \ref{fig:tidy-coefs}). Notably, complete cases underestimate the impact of room on log price. A partial residual plot (Figure \ref{fig:partial-resid}) shows there is not much variation amongst the models from the differently imputed datasets.

\begin{CodeChunk}
\begin{figure}

{\centering \includegraphics[width=1\linewidth]{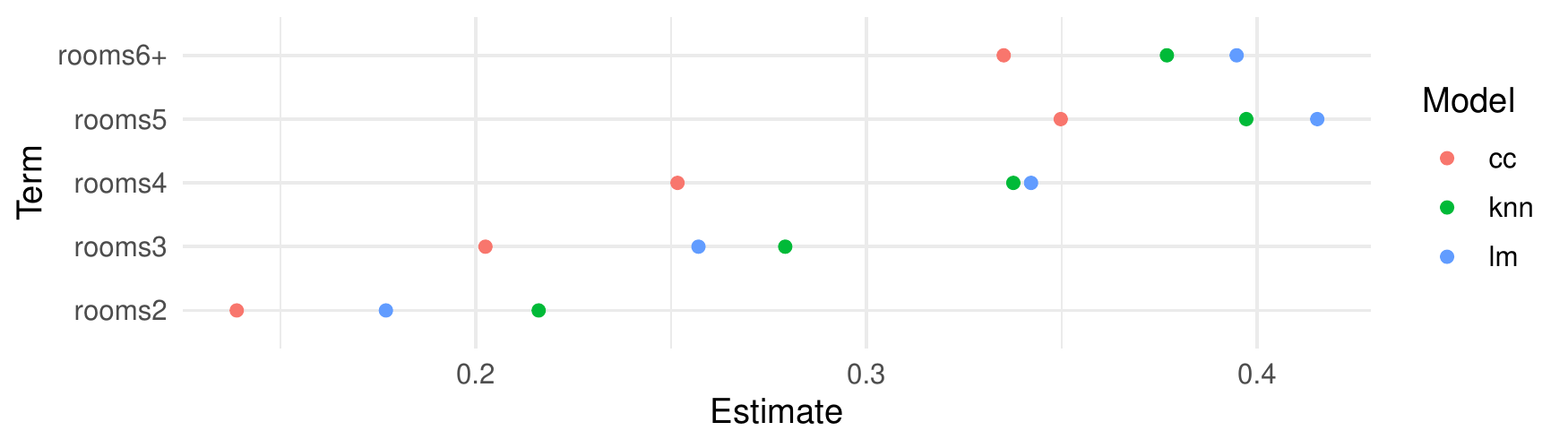} 

}

\caption[The coefficient estimate for the number of rooms varies according to the imputed dataset]{The coefficient estimate for the number of rooms varies according to the imputed dataset. Complete case dataset produced lower coefficients, compared to imputed datasets}\label{fig:tidy-coefs}
\end{figure}
\end{CodeChunk}

\begin{CodeChunk}
\begin{figure}

{\centering \includegraphics[width=1\linewidth]{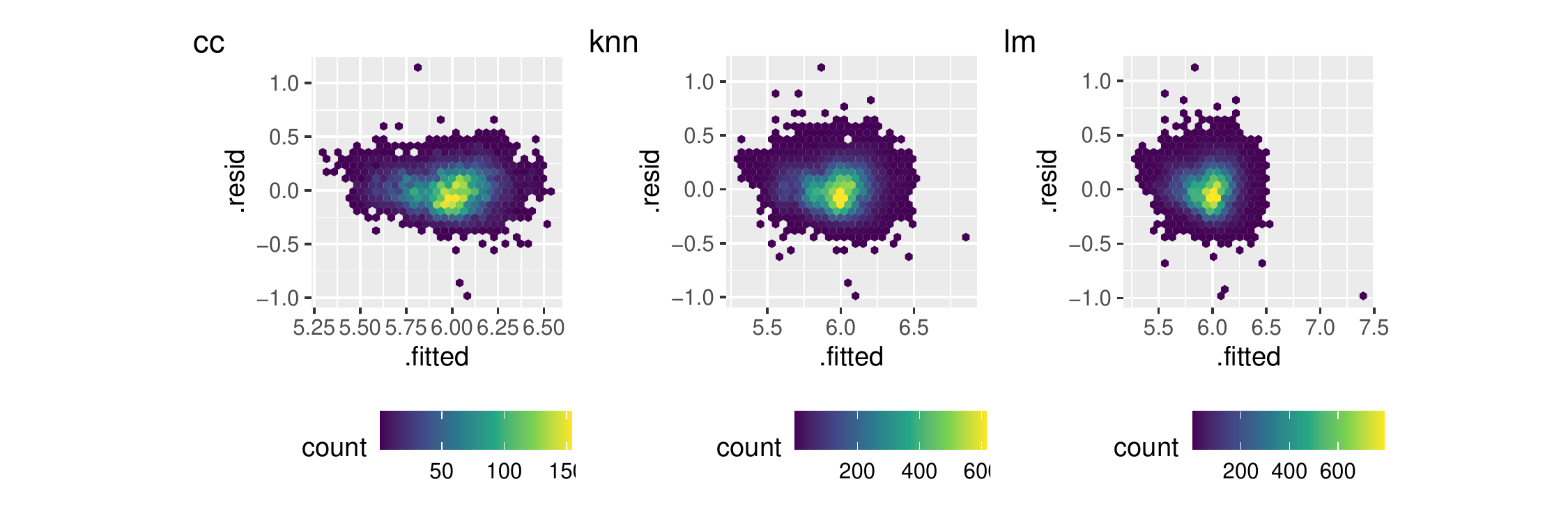} 

}

\caption[Partial residual plot for each data set, complete cases (cc), and imputed with KNN (knn) or linear model (lm)]{Partial residual plot for each data set, complete cases (cc), and imputed with KNN (knn) or linear model (lm). These are plotted as hex bins, colored according to the number of points in a given hexagon. Brighter colors mean more points. Compared to complete cases, imputed data has more points clustered around zero.}\label{fig:partial-resid}
\end{figure}
\end{CodeChunk}

\hypertarget{case-study-summary}{%
\subsection{Summary}\label{case-study-summary}}

The \texttt{naniar} and \texttt{visdat} packages implement the methods discussed in the paper, building on existing tidy tools and strike a compromise between automation and control, making analysis efficient, readable, but not overly complex. Each tool has clear intent and effects - summarising, plotting or generating data or augmenting data in some way. This not only reduces repetition and typing in an analysis, but most importantly, allows for clear expression of intent, making exploration of missing values fluent.

\hypertarget{discussion}{%
\section{Discussion}\label{discussion}}

This paper has described new methods for exploring, visualizing, and imputing missing data. The work was motivated by recent developments of tidy data, and extends them for better missing value handling. The methods have standard outputs, function arguments, and behavior. This provides consistent workflows centered around data analysis that integrate well with existing imputation methodology, visualization, and modelling.

The \emph{nabular} data structures discussed in the paper are simple by design, to promote flexibility. They could be used to create different visualizations than were shown in the paper. The analyst can use the data structures to decide on appropriate visualization for their problem. The data structures could also be used to support interactive graphics, in the manner of MANET and ggobi. Linking the plots (via linked brushing) could facilitate exploration of missingness, and could be implemented with \pkg{plotly} \citep{plotly} for added interactivity. Animating between different sets of imputed values could also be possible with packages like \pkg{gganimate} \citep{gganimate}.

Other data structures such as spatial data, time series, networks, and
longitudinal data would be supported by the inherently tabular,
\emph{nabular} data, if they are first structured as wide tidy format. Large data may need special handling, and additional features like efficient storage of purely imputed values and lazy evaluation. Special missing value codes could be improved by creating special classes, or expanding low level representation of \texttt{NA} at the source code level.

The methodology described in this paper can be used in conjunction with other approaches to understand multivariate missingness dependencies (e.g.~decision trees \citep{Tierney2015}, latent group analysis \citep{Barnett2017}, and PCA \citep{FactoMineR}). Evaluating imputed values using a testing framework like \citet{VanBuuren2012} is also supported.

The approach meshes with the dynamic nature of data analysis, allowing the analyst to go from raw data to model data in a fluid workflow.

\hypertarget{acknowledgements}{%
\section{Acknowledgements}\label{acknowledgements}}

The authors would like to thank Miles McBain, for his key contributions and discussions on the \texttt{naniar} package, in particular for helping implement \texttt{geom\_miss\_point}, and for his feedback on ideas, implementations, and names. We also thank Colin Fay for his contributions to the \texttt{naniar} package, in particular for his assistance with the \texttt{replace\_with\_na} functions. We also thank Earo Wang and Mitchell O'Hara-Wild for the many useful discussions on missing data and package development, and for their assistance with creating elegant approaches that take advantage of the tidy syntax. We would also like to thank those who contributed pull requests and discussions on the \texttt{naniar} package, in particular Jim Hester and Romain François for improving the speed of key functions, Ross Gayler for discussion on special missing values, and Luke Smith for helping \texttt{naniar} be more compliant with \texttt{ggplot2}. We would also like to thank Amelia McNamara for discussions on the paper.

\renewcommand\refname{References}
\bibliography{tidy-missing-data-paper.bib}

\end{document}